\documentclass[%
reprint,
superscriptaddress,
longbibliography,
amsmath,
amssymb,
floatfix,
prx]{revtex4-2}

\usepackage{mathtools,bbm,physics,microtype,xcolor,hyperref,amsthm}
\hypersetup{colorlinks=true,linkcolor=blue,citecolor=blue,urlcolor=blue}

\newcommand{\ad}{\operatorname{ad}}
\newcommand{\sgn}{\operatorname{sgn}}
\newcommand{\supp}{\operatorname{supp}}

\newcommand{\cH}{\mathcal{H}}
\newcommand{\cL}{\mathcal{L}}

\newcommand{\cK}{\mathcal{K}}
\newcommand{\cA}{\mathcal{A}}
\newcommand{\cB}{\mathcal{B}}
\newcommand{\cO}{\mathcal{O}}

\theoremstyle{plain}
\newtheorem{lemma}{Lemma}
\newtheorem{proposition}{Proposition}
\newtheorem{definition}{Definition}
\newtheorem{remark}{Remark}

\begin{document}

\title{Lorentz Invariant Master Equation for Quantum Systems}

\author{Pranav Vaidhyanathan}
\affiliation{
Department of Engineering Science, University of Oxford, Oxford OX1 3PJ, United Kingdom
}
\author{Gerard J. Milburn}
\affiliation{
National Quantum Computing Centre, Rutherford Appleton Laboratory, Oxfordshire, OX11 0QX, United Kingdom. 
}
\affiliation{
Sussex Centre for Quantum Technologies,
University of Sussex, Brighton BN1 9RH, United Kingdom
}

\date{\today}

\begin{abstract}
Irreversibility implies a preferred flow of time, yet special relativity denies the existence of a preferred clock. This tension has long obstructed the formulation of a relativistic master equation:
standard Markovian approximations either break Lorentz covariance, trigger catastrophic vacuum heating, or depend arbitrarily on the observer’s foliation \cite{DiosiPRD2022}. In this work, we derive a \textbf{Lorentz invariant description of irreversibility for quantum fields}. We take an approach that explicitly models the measurements required to see irreversible dynamics. Instead of evolving the system along an abstract geometric time parameter, we anchor the dynamics to a physical, relational scalar clock field. Using a relational Tomonaga-Schwinger framework, we derive a local, non-Markovian master equation that is manifestly covariant and completely positive. We show that the finite resolution of the physical clock acts as a covariant regulator, preventing the vacuum instability that plagues white-noise models. This framework demonstrates that a consistent relativistic theory of decay exists, provided the reference frame is treated as a dynamical quantum resource rather than a gauge choice.  In a gravitating context, the resulting dynamics is described by a hybrid classical--quantum (CQ) evolution that remains completely positive and trace preserving (CPTP).
\end{abstract}

\maketitle
\setcounter{page}{1}

\section{Introduction}
\label{sec:intro}

Irreversibility and decay is the ground of all experience. A physical account of irreversibility was first given in the theory of classical thermodynamics and captured by a relative ordering of states, accessible by physical interventions, and quantified by the concept of entropy. Subsequently, Boltzmann and others pioneered a description of irreversibility in terms of classical statistical mechanics. In the early twentieth century it became clear that classical statistical mechanics could not explain irreversible processes associated with black-body radiation leading to Planck's quantum postulate in 1900. It took many more decades for a consistent quantum theory of irreversibility to emerge in the work of Haken~\cite{Haken}, Louisell~\cite{Louisell} and Scully~\cite{Scully}, later formalized in the work of Gorini, Kossakowski, Lindblad, and Sudarshan assuming the Markov condition~\cite{history}. Davies~\cite{Davies} approached the description of irreversible quantum systems from the essential irreversibility of measurements. This led to the development of the ``quantum trajectories'' scheme to describe continuously measured systems in quantum optics~\cite{Carmichael,WM}. 

Attempts have been made to give a consistent Lorentz invariant description of irreversible dynamics for quantum field theory~\cite{Alicki}. Extensions to include gravity have also been proposed~\cite{Burgess}. This work attempts to generalize the GKLS master equation. Diósi~\cite{DiosiPRD2022} gave good reasons for why one should not expect Lorentz-invariant Markovian GKLS master equations for interacting relativistic quantum systems at all.

Yet there are important reasons why we need a Lorentz invariant theory of irreversibility in quantum field theory. The first reason comes from the connection between irreversibility and measurement. The LHC conducts measurements on quantum fields many times a day and the collisions involved are certainly irreversible. A master equation approach to collisions was first presented in Ref.~\cite{Borghini} and later adapted to the quantum trajectories method~\cite{BAOMAR2022108266}. A recent review examined the validity of the approximations required to use a Markov master equation~\cite{Yao} in this setting. A second reason comes from proposals to reconcile quantum theory and general relativity. Penrose~\cite{Penrose} has proposed that inevitable irreversible processes arise when quantum dynamics are dominated by gravitational interactions. While nonrelativistic treatments of the matter fields may be acceptable in lab-based experiments, relativistic extensions will be required for astrophysical and cosmological settings.

There is a necessary connection between irreversibility and time when relativistic effects are included. The ordering of physical states, at the core of irreversibility, should not depend on the reference frame of observers. Consider the example of an unstable state subject to spontaneous decay. All observers must agree on the fact of a decay event, while ascribing different spacetime coordinates to the event.    

In this paper we derive a Lorentz invariant description of irreversibility for quantum fields that avoids the problems raised by Diósi. We take an approach that explicitly models the measurements required to see irreversible dynamics.   In Section \ref{sec:rel-kinematics} we give a very general description of the relational view of time coordination by clocks. In section \ref{sec:diosi} we summarize the problems in Lorentz invariant description of irreversible dynamics based on the non relativsitic GKLS equation~\cite{history}. In section \ref{sec:PW-TS} we adopt a specific relational model based on Page--Wootters (PW) relational dynamics. The PW conditioning yields the Tomonaga--Schwinger equation (TSE). In section \ref{sec:rTS-Redfield} we partition the total system, excluding the clock, into a system plus environment and trace out the environmental degrees of freedom to obtain a near-identity positive semidefinite  map of the system state. This yields a local, non-Markovian master equation of Redfield/TCL type in the relational TS framework. In subsequent sections we investigate various mathematical features of the Master equation, the approximations required for Markov behavior, and derive the equivalent `adjoint' equations as Lorentz-covariant Heisenberg–
Langevin equation for the system field. In section \ref{sec:trajectories} we demonstrate a `quantum trajectories' representation of the non-Markov master equation. In section \ref{sec:CQ-clock} we show how describing the scalar clock as a dynamical classical field in a gravitating context,  results in a hybrid classical–quantum (CQ) evolution that remains completely positive
and trace preserving (CPTP).

\section{Relational kinematics}
\label{sec:rel-kinematics}

A clock is a device for coordinating local coincidences, for example, the change in the count on a particle detector and the angular position of the hands of a nearby clock that interacts not at all, or weakly, with the physical system of interest. This definition captures the \emph{relational} character of clock time; it refers to a relation between local, simultaneous events. Rather than the angular position of the hands of a clock, we will take the observed values, \(C\in {\mathbb R}\), of a scalar field with canonical operators \(\hat{\pi}(x), \hat{C}(x)\). The dynamics is now specified by postulating the Hamiltonian density as \(\hat{{\cal H}}_{\text{clock}}=\hat{\pi}(x)\). More complicated choices could be made. 

Consider an observer at rest with respect to a particle detector, equipped with a local clock. The particle detector is coupled to a local field in such a way that the rate of detection is given by a suitable normally ordered moment of the field. This means that the accumulating count, \(n\in {\mathbb Z}^+\), is given by a Poisson point process. The dynamics of the field is sampled by a simultaneous readout of clock value, \(C\), and the accumulating particle count \(N\). 
In this way the observer stores a measurement history \({\cal M}=\{(n_1,C_1), (n_2,C_2)\ldots (n_N, C_N)\}\). As the count is cumulative, \(n_{k+1}> n_k\). This list defines a function, \(n(C)\), that fully represents the dynamical correlations. All inertial observers will agree on the count history but they will assign different spacetime coordinates to each correlation event.

It is conventional in physics to write this in terms of a real parameterization of a pair of functions, one integer valued and one real valued, \((n(t), C(t))\). This parameterization is a mathematical convenience. In a relativistic setting we distinguish the case we have just discussed (particle detector and clock are at rest with respect to each other) by writing the real parameterization as \(\tau\) and calling it the proper time, but we stress that the physical fact of time is the relation between the events \(C\) and \(n\) and all observers agree on this. 

\begin{figure}
    \centering
    \includegraphics[scale=0.3]{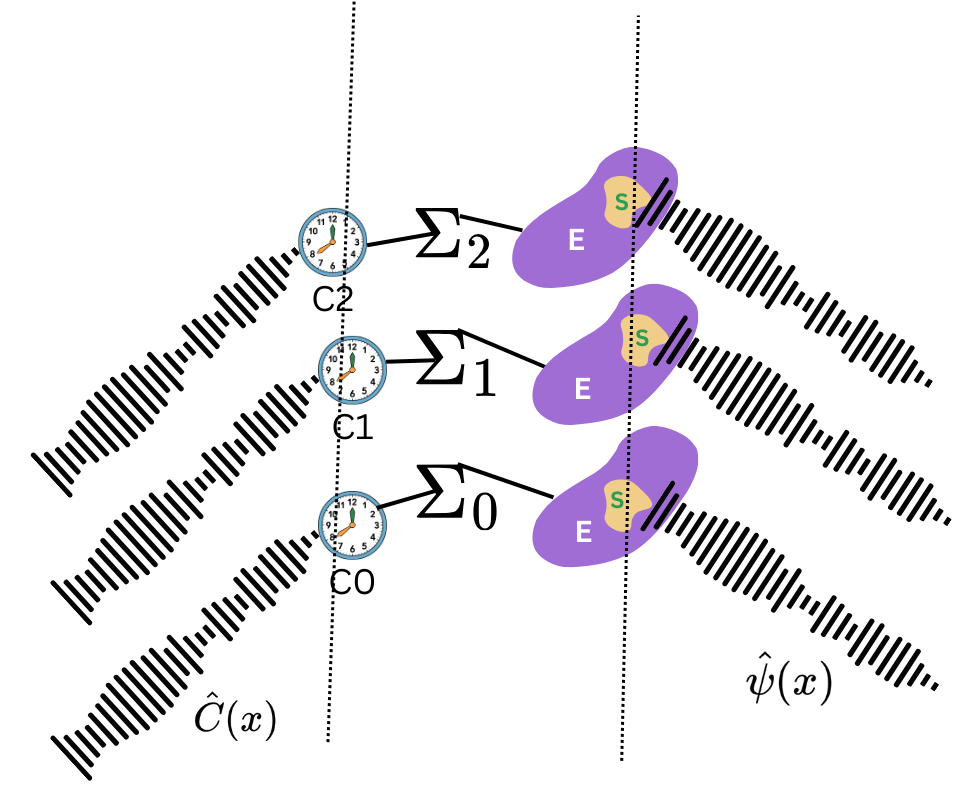}
    \caption{A scheme for relational time coordination. Two macroscopic measurement devices are at rest with respect to each other over some finite spacetime four volume, constituting a well localized laboratory. One device, `clock' (C),  makes measurements of a scalar field $C(x,t)$ with result $\textrm{C}_j$. The other device is a localized system (S),  interacting with an environment $E$ that may not be a rest with respect to C-S. Successive simultaneous measurements are made on C and S defining a  space-like hypersurface $\Sigma_j$. The measurements could be counting quanta on another field $\psi(x,t)$ interacting directly with the system. All measurements are simultaneous in the local rest frame of the clock and the system.  }
    \label{space-time}
\end{figure}

How we partition a joint quantum system is often arbitrary, but in this setting it must be unambiguous. All observers must also agree on a partition of the total quantum field into a clock and a counter. We will require that they each lie on two different geodesics in  the global spacetime. All observers will agree on this partition. It just so happens that the geodesics pass through a local neighborhood, a spacetime four volume that defines the experiment in terms of a bump (smearing) function with support on this four volume.  We  assume that this neighborhood is flat Minkowski spacetime (see Fig. \ref{space-time}).  We
must ensure that any  local decomposition used to write a GKLS form is implemented locally. In the setting of algebraic quantum field theory we require an appropriate quasi-local algebra.

Measurements produce classical random variables,  $ (C_n, M_n)$, localized to each measurement apparatus.  The statistics of the measurement results are given by a positive operator valued measure (POVM) on the state of the fields in the relevant spacetime four volume that defines the measurements. The POVM must itself be constructed from the field operators and we use Bochner positivity to ensure the resulting localized operator is positive. 

Our objective is to derive an equation to specify the state of the field in such a way as to determine the statistics of arbitrary measurement made on the system. As an example, consider the system to be a particle counter for another quantum field. We will record the history of counts, $n_k$,  at fixed increments of the clock say $C_k= k$, whose level sets \(\Sigma_k:=\{x: C(x)=k\}\) provide hypersurfaces and whose gradient fixes a unit timelike normal \(n_\mu\propto\partial_\mu C\).   We can write the measurement history in terms of the count increment on \(\Sigma_k\rightarrow \Sigma_{k+1} \)
 \begin{equation}
{\mathcal D} {\cal M}_1= \{\delta n_1, \delta n_2, \ldots ,\delta n_{N-1}\}
\end{equation}
where $\delta n_j=n_{j}-n_{j-1}$ with $n_0=0$. Stationary states are defined by ${\mathcal D} {\cal M}_1=0$, the null sequence. The count does not vary.

Suppose that the local observer is interacting with a field that is in global thermodynamic equilibrium. At the start of the experiment, the observer prepares the subsystem of clock and  counter in a far from thermal equilibrium fiducial state. For example, the clock and counter could be set so that $C=0$ and $n=0$ at the start of the experiment. Let there be many trials of the experiment starting from this fiducial state, i.e. each trial is preselected at $C=0, n=0$ on \(\Sigma_0\). It should be clear that the increments, ${\mathcal D} {\cal M}_1$  on each successive surface \(\Sigma_k \) in each trial,  will differ, even though they all start with the same fiducial state,  and the divergence between records will increase with the total length of each record. This is how irreversibility is manifest.  

We seek a fully Lorentz invariant equation that allows us to determine the statistics of the measurement record by updating the quantum state of the field at each increment of the clock,  \(\Sigma_k\rightarrow \Sigma_{k+1} \),  so that all observers agree that the incremental count record is drawn from a valid probability measure determined by a positive definite trace-class operator of trace one acting on the Hilbert space of the relevant quantum field. This equation is the relativistic GKLS equation.

\section{The relational TS framework}
\label{sec:diosi}

Diósi~\cite{DiosiPRD2022} has articulated several obstacles to constructing relativistic GKLS master equations. It is useful to summarize these points and then state how our relational Tomonaga--Schwinger framework addresses them.

First, there is a \emph{white-noise obstruction}. If one assumes inertial-time white noise, the resulting dynamics breaks Lorentz invariance. On the other hand, Lorentz invariant white noise with correlator proportional to \(\delta^{(4)}(x-y)\) leads to catastrophic vacuum heating.

Second, Diósi proposes a \emph{boost (interchange) test}. In a relativistic setting, infinitesimal evolution and infinitesimal boosts should commute. Candidate ``relativistic GKLS'' equations typically fail this test.

Third, local GKLS constructions anchored to instantaneous time slices introduce an explicit dependence on a timelike unit vector \(n^\mu\), representing the foliation normal. This dependence propagates into the dissipator and generically breaks \emph{integrability}, i.e., foliation independence of the predictions.

Finally, Diósi expresses skepticism that well-behaved, Lorentz-invariant GKLS master equations for interacting quantum fields in vacuum should exist at all.

\emph{Relational Tomonaga--Schwinger dynamics.} Our framework responds to these points as follows.

\begin{itemize}
    \item We adopt Page--Wootters (PW) relational dynamics: time is defined by a scalar clock field \(C(x)\), whose level sets \(\Sigma_\tau:=\{x: C(x)=\tau\}\) provide hypersurfaces and whose gradient fixes a unit timelike normal \(n_\mu\propto\partial_\mu C\). The foliation is therefore grounded in a physical degree of freedom with clear transformation properties, rather than a gauge choice. Conditioning on the clock leads to a Tomonaga--Schwinger equation (TSE) for the joint system--environment state. This makes covariance manifest and removes the ambiguity in choosing ``time''~\cite{PageWootters1983,Hohn2024PRD}.

    \item Tracing out the environment at the level of the TSE yields a relational TS--Redfield/TCL equation. Its kernels are built from Lorentz-scalar Wightman functions and are temporally colored, so no invariant \(\delta^{(4)}\)-noise appears and the associated vacuum instability is avoided.

    \item Under microcausality, we show that the functional derivatives of the Born/TCL\(_2\) generator commute at spacelike separation. This implies integrability (foliation independence) at that order.
    \item When the Markov limit is taken in vacuum, the generator acquires an explicit dependence on the geometric normal \(n^\mu\) through the coarse graining. This dependence cannot in general be absorbed into a covariant structure, so local vacuum GKLS dynamics is generically \emph{not} exactly Lorentz invariant. 
\end{itemize}

\subsection*{Conventions}

We work in flat Minkowski spacetime with signature \((+,-,-,-)\) and units \(\hbar=c=1\). Greek indices \(\mu,\nu \in\{0,1,2,3\}\). A hypersurface \(\Sigma_\tau\) is defined by a scalar clock field \(C(x)\) as \(\Sigma_\tau:=\{x:C(x)=\tau\}\) with timelike gradient \(\partial_\mu C\). The associated unit normal satisfies \(n_\mu \propto \partial_\mu C\) and \(n^2=1\).

\section{Relational kinematics: Page--Wootters \texorpdfstring{\(\Rightarrow\)}{⇒} Tomonaga--Schwinger}
\label{sec:PW-TS}

We now formalize the relational picture of time in quantum theory and derive the Tomonaga--Schwinger equation from the Page--Wootters construction.

Let \(\cH_S,\cH_E,\cH_C\) be the Hilbert spaces of system, environment, and clock. The timeless (Dirac) constraint enforces
\begin{equation}
  \big(\hat\pi_C(x)+\hat{\mathcal H}_{\rm tot}(x)\big)\ket{\Psi}=0
  \quad \forall x,
  \label{eq:PW-constraint}
\end{equation}
with \(\hat\pi_C(x)\) conjugate to the scalar field clock \(\hat{C}(x)\). Conditioning on a projective measurement of \(\hat{C}(x)\) with outcome \(\tau\) yields the unnormalized state on \(\Sigma_\tau\),
\begin{equation}
\varrho_{SE}[\Sigma_\tau]
  \propto \Tr_C\!\big[\ket{\tau}_C\bra{\tau}\,\ket{\Psi}\bra{\Psi}\big].
\end{equation}
This satisfies the Tomonaga--Schwinger equation (TSE)
\begin{equation}
 \frac{\delta \varrho_{SE}[\Sigma]}{\delta \Sigma(x)}
   = -i\,[\mathcal{H}_{\rm tot}(x),\varrho_{SE}[\Sigma]],
 \label{eq:TSE}
\end{equation}
which implements multi-fingered (hypersurface) evolution in QFT~\cite{PageWootters1983,Hohn2024PRD}.

The relational clock thus plays a dual role: it provides a physical foliation with a well-defined unit normal \(n_\mu\), and it supplies the conditioning that turns the timeless constraint~\eqref{eq:PW-constraint} into the dynamical TSE~\eqref{eq:TSE} for the system--environment state.

\section{From unitary TSE to local non-Markovian dynamics}
\label{sec:rTS-Redfield}

We next construct the open-system dynamics by tracing out environmental degrees of freedom at the level of the TSE. This yields a local, non-Markovian master equation of Redfield/TCL type in the relational TS framework.

Define the reduced state functional on a hypersurface \(\Sigma\) by
\begin{equation}
\rho[\Sigma]
   :=\frac{\Tr_E\varrho_{SE}[\Sigma]}
           {\Tr\,\varrho_{SE}[\Sigma]}.
\end{equation}
In the interaction picture with respect to \(\cH_0=\cH_S+\cH_E\), the TSE becomes
\begin{equation}
\frac{\delta \varrho^I[\Sigma]}{\delta \Sigma(x)}
  = -i[\mathcal{H}_{\rm int}^I(x),\varrho^I[\Sigma]].
\end{equation}
Integrating this equation over the spacetime volume \(V(\Sigma_0,\Sigma)\) between two hypersurfaces and substituting back into \(\delta\rho^I/\delta\Sigma(x)\) yields, at Born order and assuming \(\Tr_E(\mathcal{H}_{\rm int}^I\rho_E)=0\),
\begin{equation}
\begin{aligned}
\frac{\delta \rho^I[\Sigma]}{\delta \Sigma(x)}
  ={}& - \int_{V(\Sigma_0,\Sigma)}\! d^4y\,
      \Tr_E\Big(
        [\mathcal{H}_{\rm int}^I(x),\\
      &\qquad
        [\mathcal{H}_{\rm int}^I(y),
         \rho^I(y)\otimes\rho_E]]
      \Big).
\end{aligned}
\label{eq:rTS-Redfield}
\end{equation}
This is the relational TS--Redfield equation.

For a scalar local coupling of the form
\begin{equation}
\mathcal{H}_{\rm int}(x)
   =\sum_\alpha F_\alpha(x)\,
      \mathcal{J}_\alpha(x)\otimes X_\alpha(x),
\end{equation}
the Born kernels appearing in~\eqref{eq:rTS-Redfield} are Lorentz scalars, namely the Wightman functions \(G^+_{\alpha\beta}(x,y)\) and their covariant derivatives. In particular, the equation is driven by temporally colored (non-\(\delta\)-correlated) noise, and no Lorentz-invariant \(\delta^{(4)}\)-noise appears. This eliminates the vacuum-heating pathology discussed by Diósi~\cite{DiosiPRD2022}.

\paragraph{Form covariance.} Suppose that the state of the environment is Poincaré covariant and that \(\cH_{\rm int}\) is a scalar density. Let \(U(\Lambda,a)\) denote the unitary representation of the Poincaré group on the total Hilbert space. Then the rTS--Redfield equation is form covariant in the sense that
\begin{equation}
\begin{aligned}
  U(\Lambda,a)&\,
  \frac{\delta \rho[\Sigma]}{\delta \Sigma(x)}\,
  U^\dagger(\Lambda,a)\\
  &=
  \frac{\delta \rho[\Lambda\Sigma+a]}
       {\delta(\Lambda\Sigma+a)(\Lambda x + a)}.
\end{aligned}
\end{equation}
The relational TSE therefore provides a covariant starting point for open-system dynamics.

\section{TCL generator with a relational clock: positivity and stability}
\label{sec:rTCL}

To obtain a local-in-\(\Sigma\) generator in the relational TS framework, we introduce a finite clock resolution and perform a time-convolutionless (TCL) expansion.

We smear along the normal \(n^\mu(x)\) using an even, positive kernel \(w_\sigma(s)\) with characteristic width \(\sigma\). For each spacetime point \(x\) we define the displaced point
\begin{equation}
x_s:=x-s\,n(x),
\end{equation}
and the smeared correlator
\begin{equation}
C_{\alpha\beta}(x;s)
  :=F_\alpha(x)F_\beta^*(x)\,
    w_\sigma(s)\,G^+_{\alpha\beta}(x,x_s).
\end{equation}
The TCL\(_2\) coefficients are given by the full Fourier transform
\begin{align}
\kappa_{\alpha\beta}^{\rm TCL}[\Sigma;x;\omega]
  &=\int_{-\infty}^{+\infty}\!ds\,
    e^{-i\omega s}\,C_{\alpha\beta}(x;s),
\label{eq:rTCL-kappa}\\
\cH_{\rm LS}^{\rm TCL}[\Sigma;x]
  &\sim \mathrm{P}\!\!\int_{-\infty}^{+\infty}\!ds\,
    \sgn(s)\,C_{\alpha\beta}(x;s),
\label{eq:rTCL-LS}
\end{align}
where \(\cH_{\rm LS}^{\rm TCL}\) is the local Lamb-shift density and \(\mathrm{P}\) denotes the Cauchy principal value.

\subsection{Local spectral decomposition vs.\ inertial-time frequency splitting}
\label{sec:local-bohr}

A key subtlety in relativistic QFT is that the familiar \emph{global} positive/negative-frequency split of a local field
(with respect to an inertial time translation generator) is \emph{not} an operation within the local algebra; it is nonlocal and generally violates microcausality, as emphasized by Diósi~\cite{DiosiPRD2022}. Since integrability and causal locality in the Tomonaga--Schwinger setting rely on microcausality, we must ensure that any ``Bohr/secular'' decomposition used to write a GKLS form is implemented \emph{locally}, i.e.\ within the appropriate quasi-local algebra.

\begin{definition}[Clock-compatible local Bohr components]
\label{def:local-bohr}
Fix a hypersurface \(\Sigma\) and a spacetime point \(x\in\Sigma\). Let \(\cA(\cO_x)\subset\cA_{\rm loc}\) be a local algebra associated with a small neighborhood \(\cO_x\) of \(x\) (e.g.\ the causal diamond of a clock-resolution cell). Define a \emph{localized} system Hamiltonian operator
\begin{equation}
H^{\rm loc}_S(x)
  :=\int_{\Sigma}\! d^3\Sigma(z)\, \chi_x(z)\,\cH_S(z),
\label{eq:Hloc}
\end{equation}
where \(\chi_x\) is a smooth bump function supported in \(\cO_x\) and normalized so that \(\chi_x\equiv 1\) on a smaller neighborhood of \(x\). The corresponding local automorphism group on \(\cA(\cO_x)\) is
\begin{equation}
\alpha_s^{(x)}(A) := e^{i s H_S^{\rm loc}(x)} A e^{-i s H_S^{\rm loc}(x)}.
\end{equation}
For any local system operator density \(A_\alpha(x)\in \cA(\cO_x)\), we define its (possibly continuous) \emph{local Bohr components}
\begin{equation}
A_{\alpha,\omega}(x)
  := \frac{1}{2\pi}\int_{-\infty}^{+\infty}\!ds\,
      e^{i\omega s}\, \alpha_s^{(x)}\!\big(A_\alpha(x)\big),
\label{eq:local-bohr-def}
\end{equation}
in the sense of operator-valued distributions (equivalently via the spectral resolution of \(\ad_{H_S^{\rm loc}(x)}\) on \(\cA(\cO_x)\)).
\end{definition}

\begin{remark}
\label{rem:hidden-defect}
Definition~\ref{def:local-bohr} performs the spectral/Bohr decomposition \emph{inside} the local algebra \(\cA(\cO_x)\). Consequently, \(A_{\alpha,\omega}(x)\in\cA(\cO_x)\) for all \(\omega\), and the decomposition does not introduce the nonlocal inertial-time frequency splitting criticized by Diósi~\cite{DiosiPRD2022}.
\end{remark}

Denoting by \(A_{\alpha,\omega}(x)\) the clock-compatible local Bohr components (Definition~\ref{def:local-bohr}), the local generator density takes the GKLS form
\begin{equation}
\begin{aligned}
\frac{\delta \rho[\Sigma]}{\delta\Sigma(x)}
&= -\,i\big[\mathcal H_S(x)+\mathcal H_{\rm LS}^{\rm TCL}[\Sigma;x],
            \,\rho[\Sigma]\big]\\
&\quad + \sum_{\alpha,\beta,\omega}
  \kappa_{\alpha\beta}^{\rm TCL}[\Sigma;x;\omega]\\
&\qquad \times \Big(
   A_{\alpha,\omega}(x)\rho[\Sigma]A_{\beta,\omega}^\dagger(x)\\
&\qquad\qquad
   -\tfrac12\{A_{\beta,\omega}^\dagger(x)A_{\alpha,\omega}(x),
             \rho[\Sigma]\}
  \Big).
\end{aligned}
\label{eq:rTS-TCL}
\end{equation}
Equation~\eqref{eq:rTS-TCL} is local in the hypersurface and non-Markovian in origin; in general the coefficients \(\kappa_{\alpha\beta}^{\rm TCL}\) depend functionally on \(\Sigma\) through the state \(\rho[\Sigma]\).

\paragraph{Bochner route to complete positivity.}

\begin{proposition}[Complete positivity from Wightman positivity]
\label{prop:CP-Bochner}
Wightman positivity implies that the restriction of \(G^+_{\alpha\beta}(x,y)\) to timelike separations is of positive type as a function of the displacement parameter. If \(w_\sigma\) is chosen to be of positive type, then the product \(w_\sigma G^+\) is again of positive type (Bochner--Schwartz theorem). Consequently, \(\kappa_{\alpha\beta}^{\rm TCL}[\Sigma;x;\omega]\) is a positive semidefinite Kossakowski matrix density (for each \(x\) and \(\omega\)) and the generator~\eqref{eq:rTS-TCL} is completely positive and trace preserving for each hypersurface \(\Sigma\).
\end{proposition}

Thus the relational clock, together with Wightman positivity, acts as a covariant regulator that guarantees complete positivity of the effective generator. Remark~\ref{rem:hidden-defect} clarifies why this local GKLS representation does not rely on a nonlocal inertial-time frequency split.

\section{Hypersurface independence and integrability}
\label{sec:integrability}

The relational TS formalism naturally raises the question of \emph{integrability}, or hypersurface independence: does the reduced dynamics depend on the path taken in the space of hypersurfaces between two given Cauchy slices? In the unitary theory, integrability is guaranteed by microcausality. We show that the same holds for the Born/TCL\(_2\) approximation to the reduced dynamics.

Let \(\cA_{\rm loc}\) be the quasi-local algebra of observables. Assume that the local densities satisfy microcausality in the usual QFT sense, i.e.\ operators localized in spacelike separated regions commute.

\paragraph{Microcausality assumptions.}
We assume that the \emph{primitive} local densities \(\cH_S(x)\) and the local operators \(A_\alpha(x)\) used to build the interaction density are microcausal. Since the Bohr components \(A_{\alpha,\omega}(x)\) of Definition~\ref{def:local-bohr} are obtained from \(A_\alpha(x)\) by local functional calculus inside \(\cA(\cO_x)\), they inherit microcausality. Concretely, we may impose:
\begin{equation}
\begin{aligned}
[\mathcal H_S(x),\mathcal H_S(y)] &= 0,\\
[A_{\alpha,\omega}(x),A_{\beta,\omega'}(y)] &= 0,\\
\big[A_{\alpha,\omega}^\dagger(x)A_{\beta,\omega}(x),\,
&\qquad
 A_{\gamma,\omega'}^\dagger(y)A_{\delta,\omega'}(y)\big] = 0,
\end{aligned}
\quad
\text{if }(x-y)^2<0,
\label{eq:micro}
\end{equation}
with the understanding (Remark~\ref{rem:hidden-defect}) that \(A_{\alpha,\omega}(x)\) are \emph{local} components in the sense of Definition~\ref{def:local-bohr}, not inertial-time positive/negative-frequency parts.

\begin{lemma}[Integrability after Born/TCL\texorpdfstring{\(_2\)}{2}]
\label{lem:integrability}
Under the microcausality condition~\eqref{eq:micro}, the functional curl of the Born-order rTS--Redfield equation~\eqref{eq:rTS-Redfield} and of the TCL\(_2\) generator~\eqref{eq:rTS-TCL} vanishes at spacelike separation:
\begin{equation}
\left[\frac{\delta}{\delta\Sigma(x)},
      \frac{\delta}{\delta\Sigma(y)}\right]\rho[\Sigma]=0
\qquad \text{for }(x-y)^2<0.
\end{equation}
\end{lemma}

\begin{proof}
The local generator densities are sums of commutators and anticommutators of local operators at \(x\), multiplied by c-number kernels built from Lorentz scalars such as Wightman functions and their smeared variants. For spacelike separated points \(x\) and \(y\), all operator factors commute by microcausality~\eqref{eq:micro}, so the nested action of the two functional derivatives is symmetric in \(x\) and \(y\). Equal-point renormalizations (Lamb-shift terms and counterterms) are purely local and do not contribute to the curl at spacelike separation. Hence the functional curl vanishes to the order at which the expansion is performed.
\end{proof}

Therefore, evolution from an initial hypersurface \(\Sigma_0\) to a final hypersurface \(\Sigma\) is foliation independent within the Born/TCL\(_2\) approximation. Operationally, predictions are Lorentz invariant: different observers who choose different foliations between the same initial and final slices obtain the same reduced dynamics up to the order considered.

\section{The incompatibility of markovianity and vacuum covariance}
\label{sec:Markov-limit}

The TCL generator~\eqref{eq:rTS-TCL} is non-Markovian in origin. A Markovian limit can be taken under the usual assumptions of bath mixing and time-scale separation by coarse graining correlations along the normal \(n^\mu\). This procedure, however, reintroduces the geometric normal into the Kossakowski coefficients and generically destroys integrability in vacuum.

In the Markov/secular limit one obtains a local GKLS density of the form
\begin{equation}
\begin{aligned}
\frac{\delta \rho[\Sigma]}{\delta \Sigma(x)}
&\simeq -\,i\big[\,\mathcal H_S(x) + \mathcal H_{\rm LS}(x),
                 \, \rho[\Sigma]\big]\\
&\quad + \sum_{\alpha,\beta}
  \mathcal K_{\alpha\beta}(x; n(x))\\
&\qquad \times
 \Big(L_\alpha(x)\rho[\Sigma]L_\beta^\dagger(x)\\
&\qquad\qquad
   - \tfrac12\{L_\beta^\dagger(x)L_\alpha(x),\rho[\Sigma]\}\Big),
\end{aligned}
\label{eq:local-GKLS}
\end{equation}
with
\begin{equation}
\mathcal K_{\alpha\beta}(x;\omega,n(x))
  \propto
  \int_{-\infty}^{+\infty}\! ds\,
  e^{-i\omega s}\,
  G^+_{\alpha\beta}\big(x,x-sn(x)\big).
\end{equation}
The explicit dependence on \(n^\mu(x)\) arises because the coarse graining is carried out along the direction singled out by the clock. This physics is unavoidable in vacuum, where no other four-vector is available.

In general, the commutator of the local generators at two distinct spacetime points does not vanish,
\begin{equation}
\big[\cL_{\rm GKLS}(x;n(x)),\cL_{\rm GKLS}(y;n(y))\big]\neq 0,
\end{equation}
and additional terms appear in the functional curvature associated with variations of the normal. As a result, integrability fails: the reduced dynamics depends on the choice of foliation even if the underlying unitary dynamics is covariant. This is precisely the obstruction emphasized by Diósi~\cite{DiosiPRD2022}.

Thus, local GKLS master equations in vacuum are generically not fully Lorentz invariant, despite being constructed from covariant ingredients. In our framework this limitation is not a contradiction but a controlled approximation that can be improved by going back to the non-Markovian relational generator.

\subsection{Integrability tests after the Markov limit}
\label{sec:int-tests}

For practical purposes it is helpful to formulate explicit tests that a local GKLS candidate must pass in order to be foliation independent. We present two such criteria.

\paragraph{Test A: functional curl / flat-connection test.}

Define the functional curvature two-form acting on \(\rho\),
\begin{equation}
\begin{aligned}
\mathfrak{F}(x,y)[\rho]
  :={}&\left[\frac{\delta}{\delta\Sigma(x)},
           \frac{\delta}{\delta\Sigma(y)}\right]\rho\\
  ={}&\Big([\cL_x,\cL_y]+\Delta_{xy}-\Delta_{yx}\Big)\rho,
\end{aligned}
\label{eq:curl}
\end{equation}
with \(\cL_x:=\cL_{\rm GKLS}(x;n(x))\) and
\begin{equation}
\Delta_{xy}:=\big(\delta\cL_y/\delta\Sigma(x)\big)
\end{equation}
capturing the shape and normal dependence of the generator.\footnote{For strictly local GKLS generators, \(\Delta_{xy}\) reduces to variations induced by \(n^\mu\mapsto n^\mu+\delta n^\mu\), and possible extrinsic-curvature corrections if present.}

A sufficient set of conditions for \(\mathfrak{F}(x,y)\equiv 0\) is:
\begin{enumerate}
\item microcausality of local densities, so that
  \([\,\cL_x,\cL_y\,]=0\) for spacelike separated points \(x\neq y\);
\item absence of explicit normal dependence,
  \(\partial\cL_y/\partial n^\mu=0\).
\end{enumerate}
In vacuum, condition (ii) is generically violated because
\(\mathcal{K}_{\alpha\beta}(x;\omega,n)\) is obtained by spectral sampling along \(n^\mu\). By contrast, in a medium with physical four-velocity \(U^\mu(x)\) the Markovization can be performed along \(U^\mu\), and the generator depends on a physical vector field that transforms covariantly. In that case integrability can be restored, as we discuss below.

\paragraph{Test B: boost--interchange test at the generator level.}

Let \(\cB_i(\cdot)=i[K^i,\cdot]\) be the boost superoperator. For a planar evolution step along some hypersurface \(\Sigma\), interchangeability of evolution and an infinitesimal boost around the same time slice requires
\begin{equation}
\bigg[\cB_i,\; \int_\Sigma d^3x\,
  \cL_{\rm GKLS}(x;n)\bigg]=0.
\label{eq:boost-test}
\end{equation}
Local GKLS dissipators built by Fourier sampling along \(n^\mu\) generally fail~\eqref{eq:boost-test} in vacuum, in agreement with Diósi’s analysis~\cite{DiosiPRD2022, KashiwagiMatsumuraPRA2024}.

\section{Canonical GKLS forms in two regimes}
\label{sec:canonical-GKLS}

We now collect, for ease of reference, several canonical GKLS generator densities that appear in our analysis. They correspond to two physically related regimes: (A) a relational local GKLS generator that is non-Markovian in origin; and (B) a local GKLS generator in the presence of a physical medium.

\subsection{Regime A: relational local GKLS (TCL\texorpdfstring{\(_2\)}{2})}
\label{sec:R-GKLS}

To TCL\(_2\) order with the relational clock and finite resolution kernel \(w_\sigma\), the local generator density on a hypersurface \(\Sigma\) is (cf.\ Secs.~\ref{sec:rTS-Redfield} and~\ref{sec:rTCL})
\begin{equation}
\boxed{
\begin{aligned}
\frac{\delta \rho[\Sigma]}{\delta\Sigma(x)}
&= -\,i\big[\mathcal H_S(x)+\mathcal H_{\rm LS}^{\rm TCL}[\Sigma;x],
            \,\rho[\Sigma]\big]\\
&\quad + \sum_{\alpha,\beta,\omega}
   \kappa_{\alpha\beta}^{\rm TCL}[\Sigma;x;\omega]\\
&\qquad \times \Big(
   A_{\alpha,\omega}(x)\rho[\Sigma]A_{\beta,\omega}^\dagger(x)\\
&\qquad\qquad
   -\tfrac12\{A_{\beta,\omega}^\dagger(x)A_{\alpha,\omega}(x),
             \rho[\Sigma]\}
  \Big),
\end{aligned}
}
\label{eq:R-GKLS}
\end{equation}
with rates and Lamb shift
\begin{align}
\kappa_{\alpha\beta}^{\rm TCL}[\Sigma;x;\omega]
 &= \int_{-\infty}^{+\infty}\! ds\,
    e^{-i\omega s}\,
    F_\alpha(x)F_\beta^*(x)\,
 \nonumber\\
 &\qquad\times
    w_\sigma(s)\,
    G^+_{\alpha\beta}\!\big(x,x-s\,n(x)\big),
\label{eq:R-kappa}
\\[1ex]
\mathcal H_{\rm LS}^{\rm TCL}[\Sigma;x]
 &\sim
    \mathrm{P}\!\!\int_{-\infty}^{+\infty}\! ds\,
    \sgn(s)\,
    F_\alpha(x)F_\beta^*(x)\,
 \nonumber\\
 &\qquad\times
    w_\sigma(s)\,
    G^+_{\alpha\beta}\!\big(x,x-s\,n(x)\big).
\label{eq:R-LS}
\end{align}
The assumptions and guarantees are:
(i) Wightman positivity and the positive-type \(w_\sigma\) imply, via Bochner--Schwartz, that \(\kappa^{\rm TCL}\) defines a positive semidefinite Kossakowski matrix, hence Eq.~\eqref{eq:R-GKLS} is CP and TP.
(ii) Under microcausality (Lemma~\ref{lem:integrability}) and with the local Bohr components as in Definition~\ref{def:local-bohr}, the functional curl vanishes to Born/TCL\(_2\) order, so integrability holds.
(iii) Form covariance follows from the scalar nature of the kernels and the covariance of \(n^\mu\).

\subsection{Regime B: local GKLS with a physical medium}
\label{sec:U-GKLS}
When a KMS medium with four-velocity \(U^\mu(x)\) is present, the medium provides a physical time direction. The Markov coarse graining can then be taken along \(U^\mu\) rather than the geometric normal, yielding an effective local GKLS generator
\begin{equation}
\boxed{
\begin{aligned}
\frac{\delta \rho[\Sigma]}{\delta\Sigma(x)}
&\simeq -\,i\big[\mathcal H_S(x)+\mathcal H_{\rm LS}(x),
                 \,\rho[\Sigma]\big]\\
&\quad + \sum_{\alpha,\beta,\omega}
     \mathcal K^{(U)}_{\alpha\beta}(x;\omega)\\
&\qquad \times
  \Big(L_{\alpha,\omega}(x)\rho[\Sigma]L_{\beta,\omega}^\dagger(x)\\
&\qquad\qquad
   -\tfrac12\{L_{\beta,\omega}^\dagger(x)L_{\alpha,\omega}(x),
             \rho[\Sigma]\}\Big).
\end{aligned}
}
\label{eq:U-GKLS}
\end{equation}
The rates are
\begin{equation}
\begin{aligned}
\mathcal K^{(U)}_{\alpha\beta}(x;\omega)
&=\int_{-\infty}^{+\infty}\!ds\,
  e^{-i\omega s}\,F_\alpha(x)F_\beta^*(x)\,
  w_\sigma(s) \\
&\qquad\qquad\times
  G^+_{\alpha\beta}\!\big(x,x-s\,U(x)\big),\\[0.3em]
\mathcal K^{(U)}_{\alpha\beta}(x;\omega)
&=e^{-\beta(x)\omega}\,
  \mathcal K^{(U)}_{\beta\alpha}(x;-\omega),
\end{aligned}
\label{eq:U-kappa}
\end{equation}
where the second relation encodes local detailed balance in the comoving frame.

Here again complete positivity follows from Bochner positivity. Since the coarse-graining direction is now the physical vector \(U^\mu\), which transforms covariantly under Lorentz transformations, the generator has no dependence on a purely geometric normal. The functional curl condition can therefore be satisfied consistently with covariance, and integrability can be restored in this regime (see also Sec.~\ref{sec:int-tests}).

\subsection{Heisenberg--Langevin representation and covariant colored noise}
\label{sec:HL}

It is often illuminating to represent the reduced dynamics at the Heisenberg level in terms of a Langevin equation with covariant dissipation and noise kernels. This parallels the equivalent description of open systems in terms of quantum Langevin equations in quantum optics. We sketch this representation for a simple scalar model and relate it to the GKLS structures above.

\paragraph{Microscopic Langevin equation (two real scalars).}

Consider the Lagrangian density
\begin{equation}
\mathcal L
 =\tfrac12(\partial\phi_S)^2-\tfrac12 m_S^2\phi_S^2
 +\tfrac12(\partial\phi_E)^2-\tfrac12 m_E^2\phi_E^2
 - g\,\phi_S\phi_E.
\end{equation}
The Euler--Lagrange equations are
\begin{equation}
(\Box+m_S^2)\phi_S(x)=-g\,\phi_E(x),\qquad
(\Box+m_E^2)\phi_E(x)=-g\,\phi_S(x).
\label{eq:EL}
\end{equation}
Solving the environment equation with the retarded Green function \(\Delta^{R}_E\), \((\Box+m_E^2)\Delta^{R}_E=\delta^{(4)}\), \(\supp\Delta^{R}_E\subseteq\overline{V^+}\), gives
\begin{equation}
\phi_E(x)
  =\phi_E^{\rm in}(x)
   -g\!\int d^4y\ \Delta^{R}_E(x-y)\,\phi_S(y).
\label{eq:phiE-sol}
\end{equation}
Inserting~\eqref{eq:phiE-sol} into the first of Eqs.~\eqref{eq:EL} yields
\begin{equation}
\begin{aligned}
(\Box+m_S^2)\phi_S(x)
&+\int d^4y\ \Gamma(x-y)\,\phi_S(y)\\
&= \xi(x),
\end{aligned}
\label{eq:HL-exact}
\end{equation}
with
\begin{equation}
\Gamma(x-y):=g^2\,\Delta^{R}_E(x-y),\qquad
\xi(x):=-g\,\phi_E^{\rm in}(x).
\end{equation}
Equation~\eqref{eq:HL-exact} is the exact Lorentz-covariant Heisenberg--Langevin equation for the system field. The dissipation kernel \(\Gamma\) is a Lorentz-scalar distribution supported in the future light cone; the noise field \(\xi\) is a scalar operator built from the free environment field. Both transform covariantly, so Eq.~\eqref{eq:HL-exact} is form covariant.

\paragraph{Noise commutator, Hadamard correlator, and FDR.}
Let \(\Delta_E(x):=\langle[\phi_E(x),\phi_E(0)]\rangle\) be the Pauli--Jordan function (microcausal: \(\Delta_E=0\) for spacelike \(x\)), and \(G_E^{(1)}(x):=\langle\{\phi_E(x),\phi_E(0)\}\rangle\) the Hadamard function. From \(\xi=-g\phi_E^{\rm in}\) we obtain
\begin{align}
[\xi(x),\xi(y)]
 &= -\,g^2\,[\phi_E^{\rm in}(x),\phi_E^{\rm in}(y)]
  = -\,i\,g^2\,\Delta_E(x-y),
\label{eq:xi-comm}\\[0.3em]
\tfrac12\langle\{\xi(x),\xi(y)\}\rangle
 &=\tfrac{g^2}{2}\,G_E^{(1)}(x-y).
\label{eq:xi-sym}
\end{align}
In homogeneous stationary baths (in particular KMS states), passing to Fourier variables \(k^\mu\) yields the standard fluctuation--dissipation relation (FDR),
\begin{equation}
\begin{aligned}
\tilde N(k)
&:=\frac12\!\int d^4x\, e^{ik\cdot x}\langle\{\xi(x),\xi(0)\}\rangle\\
&=\frac12\coth\!\bigg(\frac{\beta\,k\!\cdot\!U}{2}\bigg)\, \tilde\rho_E(k),\\[0.3em]
\tilde\Gamma(k)
&= i\,\mathrm{p.v.}\!\int \frac{d\omega'}{\pi}\,
 \frac{\tilde\rho_E(\omega',\mathbf k)}{k^0-\omega'},
\end{aligned}
\label{eq:FDR}
\end{equation}
where \(\tilde\rho_E(k)=g^2\,\sgn(k^0)\,2\pi\,\delta(k^2-m_E^2)\) is the bath spectral density, \(U^\mu\) is the medium four-velocity (vacuum limit \(\beta\to\infty\)), and ``p.v.'' denotes principal value.

\paragraph{CCR preservation.}
Let \(\pi_S:=\partial_0\phi_S\) be the canonical momentum on some Cauchy surface \(\Sigma_0\) with the standard equal-time CCR. Denote the Pauli--Jordan function of the reduced field by \(\Delta_S(x,y):=[\phi_S(x),\phi_S(y)]\).

\begin{proposition}[CCR preserved by the Langevin equation]
\label{prop:CCR}
The commutator \(\Delta_S\) solves the homogeneous causal Volterra equation
\begin{equation}
\begin{aligned}
(\Box_x+m_S^2)\,\Delta_S(x,y)
&+\int d^4z\ \Gamma(x-z)\,\Delta_S(z,y)=0,\\
\Delta_S|_{x^0=y^0}&=0,\\
\partial_{x^0}\Delta_S|_{x^0=y^0}
&=\delta^{(3)}(\mathbf x-\mathbf y).
\end{aligned}
\label{eq:DeltaS}
\end{equation}
This equation has a unique causal solution; equivalently, the equal-time CCR are preserved for all times \(x^0\).
\end{proposition}

\begin{proof}[Sketch]
We demonstrate this by taking the commutator of Eq.~\eqref{eq:HL-exact} with \(\phi_S(y)\), using Eq.~\eqref{eq:xi-comm} and the linearity of the dynamics to show that \([\xi(x),\phi_S(y)]\) is proportional to \(\Delta_S\) smeared by \(\Delta^R_E\), and impose initial CCR on the hypersurface. Causality of \(\Gamma\) implies that the Volterra equation~\eqref{eq:DeltaS} is well posed and preserves the initial data. Microcausality of \(\Delta_E\) implies microcausality of \(\Delta_S\).
\end{proof}

\paragraph{Relational clock smearing and rTCL Langevin form.}
In the relational TS construction the coarse graining is implemented by smearing along the physical normal \(n^\mu(x)\) with an even positive-type kernel \(w_\sigma(s)\) (Sec.~\ref{sec:rTCL}). Define a smeared input noise field and a smeared retarded kernel by
\begin{align}
\Xi_\sigma(x)
 &:=\int_{-\infty}^{+\infty}\! ds\,
   w_\sigma^{1/2}(s)\,
   \xi\!\big(x-s\,n(x)\big),
\label{eq:Xi-sigma}\\[0.3em]
\Gamma_\sigma(x-y)
 &:=\int_{-\infty}^{+\infty}\! ds
   \int_{-\infty}^{+\infty}\! ds'\,
   w_\sigma^{1/2}(s)\,w_\sigma^{1/2}(s')\\
 &\qquad \times
   \Gamma\!\big(x-s\,n(x)-y+s'\,n(y)\big).
\label{eq:Gamma-sigma}
\end{align}
Then the local-in-\(\Sigma\) rTCL Heisenberg--Langevin equation reads
\begin{equation}
(\Box+m_S^2)\phi_S(x)
 + \int d^4y\ \Gamma_\sigma(x-y)\,\phi_S(y)
 =\Xi_\sigma(x),
\label{eq:HL-rTCL}
\end{equation}
with noise statistics
\begin{equation}
\begin{aligned}
[\Xi_\sigma(x),\Xi_\sigma(y)]
 &= -\,i\!\int ds\,ds'\,
    w_\sigma^{1/2}(s)\,w_\sigma^{1/2}(s')\\
 &\qquad \times
    g^2\,\Delta_E\!\big(x-s n(x)-y+s' n(y)\big),\\[0.3em]
\tfrac12\langle\{\Xi_\sigma(x),\Xi_\sigma(y)\}\rangle
 &=\tfrac{g^2}{2}\!\int ds\,ds'\,
   w_\sigma^{1/2}(s)\,w_\sigma^{1/2}(s')\\
 &\qquad \times
   G_E^{(1)}\!\big(x-s n(x)-y+s' n(y)\big).
\end{aligned}
\label{eq:Xi-stats}
\end{equation}
Wightman positivity and the positive-type \(w_\sigma\) imply that the Fourier transform of the symmetrized correlator in Eq.~\eqref{eq:Xi-stats} is positive semidefinite, matching the Kossakowski density of Sec.~\ref{sec:R-GKLS} (Bochner route to CP). Microcausality of \(\Delta_E\) implies \([\Xi_\sigma(x),\Xi_\sigma(y)]=0\) at spacelike separation, so locality is preserved.

\paragraph{Mode form and relation to the comoving GKLS generator.}
In a KMS medium with four-velocity \(U^\mu\) (Sec.~\ref{sec:U-GKLS}), align \(n^\mu=U^\mu\) and take the ideal-clock limit on bath scales (\(w_\sigma\to 1\) along \(U\)). In momentum space Eq.~\eqref{eq:HL-exact} diagonalizes on on-shell modes \(a_{\mathbf p}\), leading to the comoving quantum Langevin equation
\begin{equation}
\begin{aligned}
\partial_\tau a_{\mathbf p}(\tau)
 &=\Big(-\tfrac12\Gamma(\mathbf p)-i E_{\mathbf p}\Big)\, a_{\mathbf p}(\tau)
 +F_{\mathbf p}(\tau),\\
[F_{\mathbf p}(\tau),F_{\mathbf q}^\dagger(\tau')]
 &=\Gamma(\mathbf p)\,\delta(\tau-\tau')\, \delta^{(3)}(\mathbf p-\mathbf q),
\end{aligned}
\label{eq:mode-Langevin}
\end{equation}
with \(\tau\) the proper time along \(U^\mu\) and \(\Gamma(\mathbf p)\equiv\mathcal K^{(U)}(x;\omega=E_{\mathbf p})\) the GKLS rate from Eq.~\eqref{eq:U-GKLS}. The solution
\begin{equation}
\begin{aligned}
a_{\mathbf p}(\tau)
&=e^{-(\Gamma/2+iE_{\mathbf p})\tau}a_{\mathbf p}(0)\\
&\quad +\int_0^\tau\!d\tau'\,
  e^{-(\Gamma/2+iE_{\mathbf p})(\tau-\tau')}\, F_{\mathbf p}(\tau'),
\end{aligned}
\end{equation}
preserves the CCR,
\(
 [a_{\mathbf p}(\tau),a_{\mathbf q}^\dagger(\tau)]
 =\delta^{(3)}(\mathbf p-\mathbf q).
\)
Thus the comoving GKLS generator~\eqref{eq:U-GKLS} is precisely the Markovian limit of the rTCL Langevin equation. 

\section{Covariant quantum trajectories and stochastic dynamics}
\label{sec:trajectories}

The rTS--Redfield equation~\eqref{eq:rTS-Redfield} and its TCL$_2$ reduction~\eqref{eq:rTS-TCL}
describe the evolution of the reduced density operator $\rho[\Sigma]$ obtained after tracing out the
environment on the Page--Wootters constraint surface. It is often convenient to represent this
evolution in terms of \emph{quantum trajectories}, i.e.\ stochastic pure states $|\Psi[\Sigma]\rangle$
such that
\begin{equation}
  \rho[\Sigma] = \mathbb{E}\big[\,|\Psi[\Sigma]\rangle\langle\Psi[\Sigma]|\big],
\end{equation}
where the expectation is over realizations of some classical noise. The integrability result
of Lemma~\ref{lem:integrability} concerns the deterministic evolution of $\rho[\Sigma]$; an
unraveling $|\Psi[\Sigma]\rangle$ is not unique and need not itself be foliation independent.
Here we sketch two natural classes of trajectories compatible with the rTS dynamics.

\subsection{Non--Markovian trajectories with covariant colored noise}

At Born order the rTS--Redfield equation~\eqref{eq:rTS-Redfield} is driven by environment
Wightman functions $G^+_{\alpha\beta}(x,y)$, which are Lorentz-scalar two-point functions.
We can encode these correlators in a set of complex Gaussian scalar noise fields
$\zeta_\alpha(x)$ with
\begin{subequations}
\label{eq:zeta-noise}
\begin{align}
  \mathbb{E}[\zeta_\alpha(x)] &= 0,\\
  \mathbb{E}[\zeta_\alpha(x)\,\zeta_\beta^*(y)]
  &= G^+_{\alpha\beta}(x,y).
\end{align}
\end{subequations}
Since $G^+_{\alpha\beta}$ is a Lorentz scalar, $\zeta_\alpha(x)$ defines a covariant colored
noise field on spacetime.

By adapting non--Markovian quantum state diffusion (NMQSD) techniques to the
Tomonaga--Schwinger setting, one can write a \emph{non--Markovian stochastic TSE} for
a conditioned state $|\Psi_\zeta[\Sigma]\rangle$ in the interaction picture,
\begin{equation}
  \frac{\delta}{\delta\Sigma(x)}|\Psi_\zeta^I[\Sigma]\rangle
  = \Big(\mathcal{L}_{\mathrm{det}}(x;\Sigma)
       + \mathcal{L}_{\mathrm{stoch}}(x;\Sigma;\zeta)\Big)
    |\Psi_\zeta^I[\Sigma]\rangle,
  \label{eq:NMS-TSE}
\end{equation}
where $\mathcal{L}_{\mathrm{det}}$ contains the memory terms built from the same
kernels as~\eqref{eq:rTS-Redfield}, and $\mathcal{L}_{\mathrm{stoch}}$ couples local system
operators at $x$ to the c-number noise fields $\zeta_\alpha(x)$.
By construction,
$\mathbb{E}[|\Psi_\zeta[\Sigma]\rangle\langle\Psi_\zeta[\Sigma]|]$ satisfies the rTS--Redfield
equation to Born order. Under the microcausality assumptions of Lemma~\ref{lem:integrability},
the local generator densities at spacelike separation commute (system operators commute,
$\zeta_\alpha$ are c-numbers), so the functional curl of~\eqref{eq:NMS-TSE} vanishes to this
order and the trajectories are foliation independent.

\subsection{rTCL trajectories and hypersurface white noise}

For the local TCL$_2$ generator~\eqref{eq:rTS-TCL}, which is of GKLS form with Kossakowski
density $\kappa^{\mathrm{TCL}}_{\alpha\beta}[\Sigma;x;\omega]$, we can employ standard diffusive
unravelings. After the local secular decomposition and diagonalization of
$\kappa^{\mathrm{TCL}}$, we obtain eigenoperators $L_k(x)$ and nonnegative rates $\gamma_k(x)$.
An (unnormalized) stochastic Tomonaga--Schwinger equation for the trajectories
$|\Psi[\Sigma]\rangle$ then reads
\begin{equation}
\begin{aligned}
\frac{\delta}{\delta\Sigma(x)}|\Psi[\Sigma]\rangle
&=\Bigg(
      -i\,\mathcal{H}_{\mathrm{eff}}(x) \\
&\qquad\quad
      + \sum_k \sqrt{\gamma_k(x)}\,L_k(x)\,\xi_k(x)
    \Bigg)|\Psi[\Sigma]\rangle,
\end{aligned}
\label{eq:STSE-linear}
\end{equation}

with effective Hamiltonian density
\begin{equation}
  \mathcal{H}_{\mathrm{eff}}(x)
  = \mathcal{H}'(x)
    - \frac{i}{2}\sum_k\gamma_k(x)L_k^\dagger(x)L_k(x),
\end{equation}
where $\mathcal{H}'(x)$ includes the system Hamiltonian and the TCL Lamb shift.
The complex noise fields $\xi_k(x)$ are taken to be white on the hypersurface,
\begin{subequations}
\label{eq:hypersurface-white}
\begin{align}
  \mathbb{E}[\xi_k(x)] &= 0,\\
  \mathbb{E}[\xi_k(x)\,\xi_j^*(y)]
  &= \delta_{kj}\,\delta^{(3)}_\Sigma(x,y),
\end{align}
\end{subequations}
where $\delta^{(3)}_\Sigma(x,y)$ is the Dirac delta with respect to the induced volume
element on $\Sigma$, i.e.\ $\int_\Sigma d^3\Sigma(y)\,\delta^{(3)}_\Sigma(x,y)f(y)=f(x)$.
The noise is thus white with respect to the relational time defined by the clock field $C(x)$
(along $n^\mu$), not with respect to an arbitrary inertial time coordinate.

A standard It\^o calculation shows that the ensemble average
$\mathbb{E}[|\Psi[\Sigma]\rangle\langle\Psi[\Sigma]|]$ obeys the rTS--TCL equation
\eqref{eq:rTS-TCL}. The stochastic generator~\eqref{eq:STSE-linear} is, however, generally
foliation dependent because the covariance~\eqref{eq:hypersurface-white} involves the
hypersurface delta $\delta^{(3)}_\Sigma$.

\subsection{CSL--like interpretation and CQ clocks}

Collapse models such as CSL can be viewed as fundamental stochastic modifications of the
Schr\"odinger dynamics, whose density-matrix description takes the GKLS/SME form.
In a relativistic context, Diósi's analysis shows that attempts to drive such dynamics with
Lorentz-invariant white noise lead either to explicit frame dependence or to unphysical
vacuum heating~\cite{DiosiPRD2022}.
The rTS framework suggests instead that a viable relativistic CSL--like model should be
formulated relationally, driven by covariant colored noise (as in the non--Markovian
trajectories above) and regularized by a covariant clock kernel $w_\sigma$.

This structure is realized explicitly when the clock is promoted to a dynamical classical field
in the CQ--rTS equation~\eqref{eq:CQ-rTS}(Section~\ref{sec:CQ-clock}). There, a local classical Fokker--Planck generator
with diffusion $D_2$ drives stochastic dynamics in the classical sector, while the quantum
sector evolves under a local GKLS generator with coefficients $D_0$, constrained by the
decoherence--diffusion trade-off~\eqref{eq:CQ-tradeoff}.
Classical diffusion in the clock (and, in a gravitating extension, in the metric) thus provides
a fundamental noise source that induces irreversible decoherence/localization in the quantum
sector in a Lorentz-covariant and integrable way, without introducing $\delta^{(4)}$ quantum
white noise.

\section{Relation between regimes A and B}
\label{sec:A-to-B}

We briefly explain how the local GKLS generator with a physical medium (Regime B) arises as a limit of the relational TCL generator (Regime A).

Let \(n^\mu(x)=U^\mu(x)\), i.e., use the comoving relational time. Assume the environment is locally KMS in the comoving frame, with mixing so that its two-point functions cluster sufficiently rapidly along \(U^\mu\). If \(w_\sigma(s)\to 1\) on the bath correlation scale and the standard Born--Markov--secular scale separation holds, then
\begin{equation}
\kappa^{\rm TCL}_{\alpha\beta}[\Sigma;x;\omega]
 \xrightarrow[\ \sigma\to\infty\ ]{}\ 
 \mathcal K^{(U)}_{\alpha\beta}(x;\omega),
\end{equation}
and \(\cH_{\rm LS}^{\rm TCL}[\Sigma;x]\to \cH_{\rm LS}(x)\). Hence the time-homogeneous GKLS generator~\eqref{eq:U-GKLS} is the limit of the relational generator~\eqref{eq:R-GKLS}.

More explicitly, define
\begin{equation}
\begin{aligned}
\frac{\delta\rho}{\delta\Sigma(x)}
&= -\,i\!\left[\mathcal H_S(x)+\mathcal H^{\rm TCL}_{\rm LS}[\Sigma;x],
               \rho\right]\\
&\quad + \sum_{\alpha,\beta,\omega}
   \Big(\mathcal K^{(U)}_{\alpha\beta}(x;\omega)
       +\delta\kappa^{(\sigma)}_{\alpha\beta}[\Sigma;x;\omega]
   \Big)\\
&\qquad \times
   \Big(L_{\alpha,\omega}(x)\rho L^\dagger_{\beta,\omega}(x)\\
&\qquad\qquad
      -\tfrac12\{L^\dagger_{\beta,\omega}(x)L_{\alpha,\omega}(x),
                \rho\}\Big),
\end{aligned}
\label{eq:U-relational}
\end{equation}
where
\begin{equation}
\begin{aligned}
\delta\kappa^{(\sigma)}_{\alpha\beta}[\Sigma;x;\omega]
&=\int_{-\infty}^{\infty}\!ds\,
  e^{-i\omega s}\,F_\alpha(x)F_\beta^*(x)\\
&\qquad \times
  \big(w_\sigma(s)-1\big)\,
  G^+_{\alpha\beta}\!\big(x,x-sU(x)\big).
\end{aligned}
\label{eq:delta-kappa}
\end{equation}
For each \(\Sigma\), Eq.~\eqref{eq:U-relational} is CP, since both \(\mathcal K^{(U)}\) and \(\delta\kappa^{(\sigma)}\) are Fourier transforms of positive-type functions and their sum is again positive semidefinite. Locality and covariance are manifest. Under microcausality, integrability holds to Born/TCL\(_2\) order as in Lemma~\ref{lem:integrability}. The limit \(\sigma\to\infty\) reproduces Eq.~\eqref{eq:U-GKLS}, while finite \(\sigma\) provides a controlled memory refinement of the comoving GKLS description.

\section{Worked examples}
\label{sec:worked-examples}

We illustrate some aspects of the general framework with simple examples. A detailed field-theoretic example is included here as a concrete realization.

\subsection{Thermal/KMS media: physical time direction}

If the bath is locally KMS with four-velocity \(U^\mu(x)\) and inverse temperature \(\beta(x)\), then relational time and the medium’s rest-frame time can be aligned by choosing \(n^\mu=U^\mu\). The GKLS coarse-graining direction is then physical, and the rates satisfy local detailed balance:
\begin{equation}
\kappa_{\alpha\beta}(x;\omega)
 =e^{-\beta(x)\omega}\,
  \kappa_{\beta\alpha}(x;-\omega).
\end{equation}
Integrability is compatible with covariance because \(U^\mu\) transforms as a four-vector and carries the medium’s rest frame. In this regime the Markovian generator~\eqref{eq:U-GKLS} provides a consistent, covariant description of dissipative dynamics.

\subsection{Scalar-field model: relational rTCL generator and Markov limits}
\label{sec:scalar-example}

We now work through the scalar system--environment model sketched in Sec.~\ref{sec:HL}, in the relational rTCL framework, and connect it explicitly to the Markov limits in vacuum and in a KMS medium.

\subsubsection{Model}

Consider two real scalar fields on Minkowski spacetime:
a \emph{system} field \(\phi_S\) (mass \(m_S\)) and an \emph{environment} field \(\phi_E\) (mass \(m_E\)),
with local Lorentz-scalar interaction density
\begin{equation}
\cH_{\rm int}(x)
 =g\,\phi_S(x)\,\phi_E(x),
 \quad \rho_E=\ket{0}\!\bra{0}
 \ \text{(Minkowski vac.)}.
\label{eq:add-Hint}
\end{equation}
Time is relational: a scalar clock \(C(x)\) defines hypersurfaces \(\Sigma_\tau=\{x:C(x)=\tau\}\)
with unit timelike normal \(n^\mu(x)\propto\partial^\mu C(x)\).
Conditioning on \(C\) yields the Tomonaga--Schwinger equation for the joint state on \(\Sigma\) as in Sec.~\ref{sec:PW-TS}.

\subsubsection{Clock smearing and local rTCL\texorpdfstring{\(_2\)}{2} generator}

Advance \(x\) along the clock direction by \(x_s:=x-s\,n(x)\) and introduce an even, positive-type kernel
\(w_\sigma(s)\) (finite clock resolution). Define the smeared scalar kernel
\begin{equation}
\begin{aligned}
C(x;s)
&:=g^2\,w_\sigma(s)\,G_E^+\!\big(x,x_s\big)\\
&= g^2\,w_\sigma(s)\!\int\!\frac{d^3\mathbf k}{(2\pi)^3\,2E_{\mathbf k}}\,
   e^{-i s\,k\cdot n(x)}.
\end{aligned}
\label{eq:add-C}
\end{equation}

\emph{Local Bohr components.} To remain consistent with microcausality, \(A_\omega(x)\) should be understood as the \emph{clock-compatible local} components of Definition~\ref{def:local-bohr} applied to a locally smeared system observable (e.g.\ \(\phi_S\) smeared in a small neighborhood \(\cO_x\)), rather than a global inertial-time frequency split of \(\phi_S\).

At TCL\(_2\) order (specializing Eqs.~\eqref{eq:R-GKLS}--\eqref{eq:R-LS}),
\begin{equation}
\begin{aligned}
\frac{\delta \rho[\Sigma]}{\delta\Sigma(x)}
&= -\,i\big[\cH_S(x)+\cH_{\rm LS}^{\rm TCL}[\Sigma;x],
            \,\rho[\Sigma]\big]\\
&\quad + \int_{-\infty}^{+\infty}\!d\omega\,
   \kappa^{\rm TCL}(x;\omega)\\
&\qquad \times
   \Big(A_\omega(x)\rho[\Sigma]A_\omega^\dagger(x)\\
&\qquad\qquad
   -\tfrac12\{A_\omega^\dagger(x)A_\omega(x),\rho[\Sigma]\}
   \Big),
\end{aligned}
\label{eq:add-rTCL}
\end{equation}
with coefficients
\begin{equation}
\begin{aligned}
\kappa^{\rm TCL}(x;\omega)
 &=\int_{-\infty}^{+\infty}\!ds\,
   e^{-i\omega s}\,C(x;s),\\[0.3em]
\cH_{\rm LS}^{\rm TCL}[\Sigma;x]
 &\sim \mathrm{P}\!\!\int_{-\infty}^{+\infty}\!ds\,
   \sgn(s)\,C(x;s).
\end{aligned}
\label{eq:add-kappa-HLS-def}
\end{equation}
As in the general discussion, \(s\mapsto G_E^+(x,x-sn)\) is positive-type along any timelike \(n^\mu\), and choosing \(w_\sigma\) positive-type makes \(w_\sigma G_E^+\) positive-type. Thus \(\kappa^{\rm TCL}(x;\omega)\) defines a positive semidefinite Kossakowski density.

\subsubsection{Gaussian clock kernel}

Choose
\begin{equation}
w_\sigma(s)=\exp\!\big(-s^2/(2\sigma^2)\big),
\end{equation}
whose Fourier transform is
\begin{equation}
\widehat w_\sigma(\Omega)
 =\sqrt{2\pi}\,\sigma\,
   e^{-\tfrac12\sigma^2\Omega^2}.
\end{equation}
Exchanging the \(s\) and \(\mathbf k\) integrals,
\begin{equation}
\boxed{
\begin{aligned}
\kappa^{\rm TCL}(x;\omega)
&=
g^2\!\int\!\frac{d^3\mathbf k}{(2\pi)^3\,2E_{\mathbf k}}\,
 \widehat w_\sigma\!\big(\omega+k\!\cdot\!n(x)\big)\\
&=
g^2\,\sqrt{2\pi}\,\sigma\!\int\!\frac{d^3\mathbf k}{(2\pi)^3\,2E_{\mathbf k}}\,
 e^{-\tfrac{\sigma^2}{2}\,(\omega+k\!\cdot\!n(x))^2}.
\end{aligned}
}
\label{eq:add-kappa-Gauss}
\end{equation}
This expression is manifestly nonnegative and scalar at \(x\) and the associated $\omega + k \cdot n(x)$ provides the Doppler shift of the bath modes relative to the local clock frame.

\paragraph{Lamb shift.}

We need the odd transform
\begin{equation}
\begin{aligned}
I(\Omega)
&:=\int_{-\infty}^{+\infty}\!ds\,
   \sgn(s)\,e^{-\tfrac{s^2}{2\sigma^2}}\,e^{-i\Omega s}\\
&= -\,2i\!\int_0^\infty\!ds\,
   e^{-\tfrac{s^2}{2\sigma^2}}\sin(\Omega s).
\end{aligned}
\end{equation}
Using
\begin{equation}
\int_0^\infty e^{-a s^2}\sin(bs)\,ds
 =\frac12\sqrt{\frac{\pi}{a}}\,
   e^{-\tfrac{b^2}{4a}}\,
   \mathrm{erfi}\!\bigg(\frac{b}{2\sqrt{a}}\bigg)
\end{equation}
with \(a=1/(2\sigma^2)\) gives the equivalent closed forms
\begin{equation}
\boxed{
\begin{aligned}
I(\Omega)
&= -\,i\,\sqrt{2\pi}\,\sigma\,
    e^{-\tfrac12\sigma^2\Omega^2}\,
    \mathrm{erfi}\!\Big(\frac{\sigma\Omega}{\sqrt{2}}\Big)\\
&= -\,i\,2\sqrt{2}\,\sigma\,
    D\!\Big(\frac{\sigma\Omega}{\sqrt{2}}\Big),
\end{aligned}
}
\label{eq:add-odd-FT}
\end{equation}
where \(D\) is the Dawson function, related to the imaginary error function by
\(e^{-z^2}\mathrm{erfi}(z)=(2/\sqrt{\pi})\,D(z)\). Therefore
\begin{equation}
\boxed{
\begin{aligned}
\cH_{\rm LS}^{\rm TCL}[\Sigma;x]
 \sim{}&
-\,i\,2\sqrt{2}\,g^2\,\sigma
\!\int\!\frac{d^3\mathbf k}{(2\pi)^3\,2E_{\mathbf k}}\,
 D\!\Big(\frac{\sigma\,k\!\cdot\!n(x)}{\sqrt{2}}\Big)\\
&\times
\big[\text{local quadratic monomial in }\phi_S(x)\big].
\end{aligned}
}
\label{eq:add-HLS-Dawson}
\end{equation}
The UV tail \(D(z)\sim 1/(2z)\) implies only local counterterms (mass and field renormalization of \(\phi_S\)),
which do not affect CP, covariance, or integrability.

\subsubsection{Vacuum Markov limit and explicit rates}

In the ideal-clock limit \(\widehat w_\sigma(\Omega)\to 2\pi\delta(\Omega)\),
\begin{equation}
\boxed{
\begin{aligned}
\kappa^{\rm Markov}(x;\omega)
&= g^2\!\int\!\frac{d^3\mathbf k}{(2\pi)^3\,2E_{\mathbf k}}\,
  2\pi\,\delta\!\big(\omega+k\!\cdot\!n(x)\big).
\end{aligned}
}
\label{eq:add-kappa-Markov}
\end{equation}
In the rest frame of \(n^\mu=(1,0,0,0)\),
\(\delta(\omega+k\!\cdot\!n)=\delta(\omega+E_{\mathbf k})\) with \(E_{\mathbf k}\ge m_E\).
A standard change of variables gives
\begin{equation}
\boxed{
\kappa^{\rm Markov}(\omega)
  =\frac{g^2}{2\pi}\,\sqrt{\omega^2-m_E^2}\;
   \Theta(-\omega-m_E),
}
\label{eq:add-kappa-Markov-closed}
\end{equation}
so no vacuum excitation (no heating) occurs, but the explicit \(n^\mu\) in Eq.~\eqref{eq:add-kappa-Markov}
reintroduces the foliation/boost obstruction for local Markov vacuum GKLS generators, exactly as discussed in Sec.~\ref{sec:Markov-limit}.

\subsubsection{KMS/comoving Markov regime as a limit of rTCL}

For a local KMS state with four-velocity \(U^\mu(x)\) and inverse temperature \(\beta(x)\),
\begin{equation}
\begin{aligned}
G^{+}_{E,\rm KMS}(x,y)
 &=\int\!\frac{d^3\mathbf k}{(2\pi)^3\,2E_{\mathbf k}}\,
 \big[(1+n_B(k\!\cdot\!U))e^{-ik\cdot(x-y)}\\
 &\qquad\quad +n_B(k\!\cdot\!U)e^{+ik\cdot(x-y)}\big],
\end{aligned}
\label{eq:add-KMS-Wightman}
\end{equation}
with \(n_B(E)=(e^{\beta E}-1)^{-1}\).
Align the clock with the medium (\(n^\mu=U^\mu\)). For finite \(\sigma\),
\begin{equation}
\begin{aligned}
\kappa^{\rm TCL}_{\rm KMS}(x;\omega)
&=
g^2\!\int\!\frac{d^3\mathbf k}{(2\pi)^3\,2E_{\mathbf k}}\,
\Big[(1+n_B)\,\widehat w_\sigma(\omega+k\!\cdot\!U)\\
&\qquad\qquad\qquad\qquad
  +(n_B)\,\widehat w_\sigma(\omega-k\!\cdot\!U)\Big].
\end{aligned}
\label{eq:add-kappa-KMS-finite}
\end{equation}
If \(\widehat w_\sigma\to 2\pi\delta\) on the bath correlation scale (ideal clock) and \(n^\mu=U^\mu\),
\begin{equation}
\boxed{
\begin{aligned}
\cK^{(U)}(x;\omega)
&=
g^2\!\int\!\frac{d^3\mathbf k}{(2\pi)^3\,2E_{\mathbf k}}\,
\Big[(1+n_B)\,2\pi\,\delta(\omega+k\!\cdot\!U)\\
&\qquad\qquad\qquad\qquad
 +(n_B)\,2\pi\,\delta(\omega-k\!\cdot\!U)\Big],
\end{aligned}
}
\label{eq:add-KMS-Markov}
\end{equation}
which satisfies detailed balance \(\cK^{(U)}(\omega)=e^{-\beta\omega}\cK^{(U)}(-\omega)\) in the comoving frame and reproduces the comoving GKLS generator~\eqref{eq:U-GKLS}. In this regime integrability is compatible with covariance because the coarse-graining direction is the physical \(U^\mu\).

\subsubsection{Clock readout and positive-type kernels}

As an explicit example of a positive-type clock kernel arising from a physical readout, consider a coherent-state (displacement) measurement on a clock mode,
\(\ket{\alpha(\tau)}=D\!\big(R\,e^{-i\omega_C\tau}\big)\ket{0}\),
with overlap
\begin{equation}
g(s)=\braket{\alpha(\tau)}{\alpha(\tau+s)}
 =\exp[-R^2(1-e^{i\omega_C s})].
\end{equation}
Its even part is a convex mixture of cosines,
\begin{equation}
w_\sigma(s)
 =e^{-R^2}\sum_{n=0}^{\infty}\frac{(R^2)^n}{n!}\,
   \cos(n\,\omega_C s),
\end{equation}
which is positive-type. For small \(s\),
\(w_\sigma(s)\approx e^{-\tfrac12 R^2\omega_C^2 s^2}\), so it behaves as a Gaussian kernel with width \(\sigma\sim 1/(R\omega_C)\). Such kernels can be used in place of the Gaussian in Eq.~\eqref{eq:add-kappa-Gauss} with the same CP, covariance, and integrability guarantees.

\section{Mathematical well-posedness}
\label{sec:well-posedness}

The local densities appearing in the TCL generator~\eqref{eq:R-GKLS} are typically unbounded operators on the underlying Hilbert space. On the quasi-local algebra of observables, the TCL density can be shown to be closable on natural dense domains. Under standard regularity and boundedness conditions on the Kossakowski kernels, conservativity and complete positivity of the effective GKLS semigroups generated by these densities follow from the criteria of Chebotarev and Fagnola~\cite{ChebotarevFagnola1998}, applied pointwise to the densities. This provides mathematical control over the reduced dynamics beyond the formal perturbation expansion.

\section{Dynamical (gravitating) clock via classical--quantum dynamics}
\label{sec:CQ-clock}

So far the clock field \(C(x)\) has been treated as an idealized reference system. We now outline how to promote the scalar clock to a dynamical classical field and, in a gravitating context, to part of the classical geometric sector. The resulting dynamics is described by a hybrid classical--quantum (CQ) evolution that remains completely positive and trace preserving (CPTP).

Following Oppenheim’s framework~\cite{OppenheimPRX2023,OppenheimNC2023,OppenheimPRD2024}, let \(z(x)\) denote a set of classical fields, including at minimum \(C,\pi_C\) and possibly metric degrees of freedom, and let \(\Upsilon[\Sigma;z]\) denote the hybrid state, an operator-valued density over the classical field phase space. A local advance of \(\Sigma\) at \(x\) satisfies the CQ--rTS equation
\begin{equation}
\frac{\delta \Upsilon}{\delta\Sigma(x)}
= -\,i\,[\mathcal H_S(x),\Upsilon]
\;+\;\mathbb L^{\rm cl}_{z}(x)\Upsilon
\;+\;\mathcal L_{\rm CQ}(x)\Upsilon,
\label{eq:CQ-rTS}
\end{equation}
where \(\mathbb L^{\rm cl}_{z}\) is a local classical Fokker--Planck generator acting on \(z\) with drift \(D_{1}^{\rm br}\) (backaction) and diffusion \(D_2\),
\begin{equation}
\begin{aligned}
\mathbb L^{\rm cl}_{z}(x)\Upsilon
&= -\partial_{z^a(x)}\!\big[\,D_{1,a}^{\rm br}(x;z)\,\Upsilon\,\big]\\
&\quad +\tfrac12\,\partial_{z^a(x)}\partial_{z^b(x)}\!
   \big[\,D_{2}^{ab}(x;z)\,\Upsilon\,\big],
\end{aligned}
\end{equation}
and \(\mathcal L_{\rm CQ}(x)\) is a GKLS generator acting on the quantum system:
\begin{equation}
\begin{aligned}
\mathcal L_{\rm CQ}(x)\Upsilon
 &= \sum_{\mu,\nu} D_{0,\mu\nu}(x;z)\Big(
     L_\mu(x;z)\,\Upsilon\,L_\nu^\dagger(x;z)\\
 &\qquad\qquad\qquad
     -\tfrac12\{L_\nu^\dagger(x;z)L_\mu(x;z),\Upsilon\}
   \Big).
\end{aligned}
\end{equation}
Complete positivity and consistency of the CQ coupling impose a decoherence--diffusion trade-off~\cite[Eq.~(23)]{OppenheimNC2023}:
\begin{equation}
\boxed{
\begin{aligned}
 D_{1}^{\rm br}\, D_{0}^{\dagger-1}\, (D_{1}^{\rm br})^\dagger
 \;&\preceq\; 2\,D_{2}
\end{aligned}
}
\label{eq:CQ-tradeoff}
\end{equation}
(in matrix-kernel sense for fields; \(D_{0}^{\dagger-1}\) is the Moore--Penrose inverse). Equivalently, there are observational trade-offs between drift, diffusion, and Lindbladian strengths~\cite[Eqs.~(20),(26),(27)]{OppenheimNC2023}.  A similar model arises in the most general formulation of measurement based control of quantum systems\cite{PhysRevLett.129.050401}.
\vskip 0.2 truecm

\paragraph{Relational foliation preserved.}

The foliation remains relational: the hypersurfaces are defined by the level sets of the classical clock field, \(\Sigma_\tau=\{x:C(x)=\tau\}\), and \(n_\mu\propto\partial_\mu C\). Thus \(n_\mu\) is a physical vector sourced by a scalar, and covariance of the construction is manifest. Form covariance of Eq.~\eqref{eq:CQ-rTS} follows since the classical tensors and the Lindblad kernels transform appropriately as scalars or tensors.

\paragraph{Integrability in the CQ setting.}

Assuming (i) microcausality of the local quantum densities and (ii) local classical kernels \(D_0(x,y)\sim\delta^{(3)}(\mathbf x-\mathbf y)\) and \(D_2(x,y)\sim\delta^{(3)}(\mathbf x-\mathbf y)\) (as expected for diffeomorphism invariance~\cite[around Eqs.~(30)--(31)]{OppenheimNC2023}), the functional curl of Eq.~\eqref{eq:CQ-rTS} vanishes at spacelike separation to Born/TCL\(_2\) order. The classical part acts by c-number differential operators on independent fields \(z(x)\), hence commutes at spacelike separation when the kernels are local. This extends Lemma~\ref{lem:integrability} to the CQ case. Markovizing along \(n^\mu\) again reintroduces explicit normal dependence, and with a stochastic \(n^\mu\) it augments the shape term \(\Delta_{xy}\) in the functional curl; in vacuum, local GKLS dynamics remains generically nonintegrable~\cite{DiosiPRD2022}.

\paragraph{White-noise issue.}

The CQ construction avoids the relativistic quantum white-noise pathology discussed by Diósi: the diffusion is carried by the classical sector and constrained by the trade-off~\eqref{eq:CQ-tradeoff}, while the quantum sector obeys a CPTP GKLS structure derived from bona fide classical--quantum kernels rather than \(\delta^{(4)}\)-correlated quantum noise.

\section{Conclusions}
\label{sec:conclusions}

We have developed a relational Tomonaga--Schwinger framework for open quantum systems in relativistic quantum field theory, in which time is defined by a scalar clock field. The clock’s level sets provide physical hypersurfaces with unit normal \(n^\mu\propto\partial^\mu C\), removing the ambiguity associated with an abstract choice of foliation.

Starting from the Page--Wootters constraint, conditioning on the clock yields the Tomonaga--Schwinger equation for the joint system--environment state. Tracing out the environment at the level of hypersurface functionals leads to a relational TS--Redfield/TCL master equation driven by Lorentz-scalar Wightman functions. Introducing a finite clock resolution and performing a TCL\(_2\) expansion yields a local generator with GKLS form, whose Kossakowski matrix is guaranteed to be positive semidefinite by Wightman positivity and Bochner’s theorem. Under microcausality, the functional curl of this generator vanishes at spacelike separation, so the dynamics is foliation independent to the order considered.

Taking a Markov limit in vacuum reintroduces explicit dependence on the geometric normal \(n^\mu\), and we showed that local GKLS generators in this regime generically fail integrability and the boost--interchange test. This embraces Diósi’s critique: one should not expect local, vacuum GKLS master equations to be exactly Lorentz invariant. By contrast, in the presence of a medium with four-velocity \(U^\mu\) the coarse graining can be carried out along \(U^\mu\), and a comoving Markovian GKLS generator can be constructed that is compatible with covariance and integrability. Finally, if locality in position space is relaxed and one works with on-shell modes and invariant measure, Poincaré-covariant and microcausal GKLS semigroups exist, as exemplified by the nonlocal generator of Ref.~\cite{KashiwagiMatsumuraPRA2024}.

The relational clock thus plays a central conceptual and technical role: it grounds the notion of time in a physical degree of freedom, acts as a covariant regulator ensuring complete positivity, and clarifies the status of Markovian approximations in relativistic open quantum dynamics. Extensions of this framework to dynamical clocks and gravitating settings can be formulated within a classical--quantum hybrid theory that remains CPTP and diffeomorphism invariant.

\bigskip

\bibliography{bibliography}

\begin{thebibliography}{23}%
\makeatletter
\providecommand \@ifxundefined [1]{%
 \@ifx{#1\undefined}
}%
\providecommand \@ifnum [1]{%
 \ifnum #1\expandafter \@firstoftwo
 \else \expandafter \@secondoftwo
 \fi
}%
\providecommand \@ifx [1]{%
 \ifx #1\expandafter \@firstoftwo
 \else \expandafter \@secondoftwo
 \fi
}%
\providecommand \natexlab [1]{#1}%
\providecommand \enquote  [1]{``#1''}%
\providecommand \bibnamefont  [1]{#1}%
\providecommand \bibfnamefont [1]{#1}%
\providecommand \citenamefont [1]{#1}%
\providecommand \href@noop [0]{\@secondoftwo}%
\providecommand \href [0]{\begingroup \@sanitize@url \@href}%
\providecommand \@href[1]{\@@startlink{#1}\@@href}%
\providecommand \@@href[1]{\endgroup#1\@@endlink}%
\providecommand \@sanitize@url [0]{\catcode `\\12\catcode `\$12\catcode `\&12\catcode `\#12\catcode `\^12\catcode `\_12\catcode `\%12\relax}%
\providecommand \@@startlink[1]{}%
\providecommand \@@endlink[0]{}%
\providecommand \url  [0]{\begingroup\@sanitize@url \@url }%
\providecommand \@url [1]{\endgroup\@href {#1}{\urlprefix }}%
\providecommand \urlprefix  [0]{URL }%
\providecommand \Eprint [0]{\href }%
\providecommand \doibase [0]{https://doi.org/}%
\providecommand \selectlanguage [0]{\@gobble}%
\providecommand \bibinfo  [0]{\@secondoftwo}%
\providecommand \bibfield  [0]{\@secondoftwo}%
\providecommand \translation [1]{[#1]}%
\providecommand \BibitemOpen [0]{}%
\providecommand \bibitemStop [0]{}%
\providecommand \bibitemNoStop [0]{.\EOS\space}%
\providecommand \EOS [0]{\spacefactor3000\relax}%
\providecommand \BibitemShut  [1]{\csname bibitem#1\endcsname}%
\let\auto@bib@innerbib\@empty
\bibitem [{\citenamefont {Di{\'o}si}(2022)}]{DiosiPRD2022}%
  \BibitemOpen
  \bibfield  {author} {\bibinfo {author} {\bibfnamefont {L.}~\bibnamefont {Di{\'o}si}},\ }\bibfield  {title} {\bibinfo {title} {Is there a relativistic {GKLS} master equation?},\ }\href@noop {} {\bibfield  {journal} {\bibinfo  {journal} {Physical Review D}\ }\textbf {\bibinfo {volume} {106}},\ \bibinfo {pages} {L051901} (\bibinfo {year} {2022})}\BibitemShut {NoStop}%
\bibitem [{\citenamefont {Haken}\ \emph {et~al.}(1967)\citenamefont {Haken}, \citenamefont {Risken},\ and\ \citenamefont {Weidlich}}]{Haken}%
  \BibitemOpen
  \bibfield  {author} {\bibinfo {author} {\bibfnamefont {H.}~\bibnamefont {Haken}}, \bibinfo {author} {\bibfnamefont {H.}~\bibnamefont {Risken}},\ and\ \bibinfo {author} {\bibfnamefont {W.}~\bibnamefont {Weidlich}},\ }\bibfield  {title} {\bibinfo {title} {Quantum mechanical solutions of the laser master equation},\ }\href@noop {} {\bibfield  {journal} {\bibinfo  {journal} {Zeitschrift f{\"u}r Physik}\ }\textbf {\bibinfo {volume} {206}},\ \bibinfo {pages} {355} (\bibinfo {year} {1967})}\BibitemShut {NoStop}%
\bibitem [{\citenamefont {Louisell}(1973)}]{Louisell}%
  \BibitemOpen
  \bibfield  {author} {\bibinfo {author} {\bibfnamefont {W.~H.}\ \bibnamefont {Louisell}},\ }\href@noop {} {\emph {\bibinfo {title} {Quantum Statistical Properties of Radiation}}}\ (\bibinfo  {publisher} {Wiley},\ \bibinfo {address} {New York},\ \bibinfo {year} {1973})\BibitemShut {NoStop}%
\bibitem [{\citenamefont {Scully}\ and\ \citenamefont {Lamb}(1967)}]{Scully}%
  \BibitemOpen
  \bibfield  {author} {\bibinfo {author} {\bibfnamefont {M.~O.}\ \bibnamefont {Scully}}\ and\ \bibinfo {author} {\bibfnamefont {W.~E.}\ \bibnamefont {Lamb}},\ }\bibfield  {title} {\bibinfo {title} {Quantum theory of an optical maser},\ }\href@noop {} {\bibfield  {journal} {\bibinfo  {journal} {Physical Review}\ }\textbf {\bibinfo {volume} {159}},\ \bibinfo {pages} {208} (\bibinfo {year} {1967})}\BibitemShut {NoStop}%
\bibitem [{\citenamefont {Chru{\'s}ci{\'n}ski}\ and\ \citenamefont {Pascazio}(2017)}]{history}%
  \BibitemOpen
  \bibfield  {author} {\bibinfo {author} {\bibfnamefont {D.}~\bibnamefont {Chru{\'s}ci{\'n}ski}}\ and\ \bibinfo {author} {\bibfnamefont {S.}~\bibnamefont {Pascazio}},\ }\bibfield  {title} {\bibinfo {title} {A brief history of the {GKLS} equation},\ }\href@noop {} {\bibfield  {journal} {\bibinfo  {journal} {Open Systems \& Information Dynamics}\ }\textbf {\bibinfo {volume} {24}},\ \bibinfo {pages} {1740001} (\bibinfo {year} {2017})}\BibitemShut {NoStop}%
\bibitem [{\citenamefont {Davies}(1976)}]{Davies}%
  \BibitemOpen
  \bibfield  {author} {\bibinfo {author} {\bibfnamefont {E.~B.}\ \bibnamefont {Davies}},\ }\href@noop {} {\emph {\bibinfo {title} {Quantum Theory of Open Systems}}}\ (\bibinfo  {publisher} {Academic Press},\ \bibinfo {year} {1976})\BibitemShut {NoStop}%
\bibitem [{\citenamefont {Carmichael}(1993)}]{Carmichael}%
  \BibitemOpen
  \bibfield  {author} {\bibinfo {author} {\bibfnamefont {H.~J.}\ \bibnamefont {Carmichael}},\ }\href@noop {} {\emph {\bibinfo {title} {An Open Systems Approach to Quantum Optics}}},\ \bibinfo {series} {Lecture Notes in Physics Monographs}, Vol.~\bibinfo {volume} {18}\ (\bibinfo  {publisher} {Springer},\ \bibinfo {year} {1993})\BibitemShut {NoStop}%
\bibitem [{\citenamefont {Wiseman}\ and\ \citenamefont {Milburn}(2009)}]{WM}%
  \BibitemOpen
  \bibfield  {author} {\bibinfo {author} {\bibfnamefont {H.~M.}\ \bibnamefont {Wiseman}}\ and\ \bibinfo {author} {\bibfnamefont {G.~J.}\ \bibnamefont {Milburn}},\ }\href@noop {} {\emph {\bibinfo {title} {Quantum Measurement and Control}}}\ (\bibinfo  {publisher} {Cambridge University Press},\ \bibinfo {year} {2009})\BibitemShut {NoStop}%
\bibitem [{\citenamefont {Alicki}\ \emph {et~al.}(1984)\citenamefont {Alicki}, \citenamefont {Fannes},\ and\ \citenamefont {Verbeure}}]{Alicki}%
  \BibitemOpen
  \bibfield  {author} {\bibinfo {author} {\bibfnamefont {R.}~\bibnamefont {Alicki}}, \bibinfo {author} {\bibfnamefont {M.}~\bibnamefont {Fannes}},\ and\ \bibinfo {author} {\bibfnamefont {A.}~\bibnamefont {Verbeure}},\ }\bibfield  {title} {\bibinfo {title} {Relativistic quantum markovian dynamics of scalar fields},\ }\href@noop {} {\bibfield  {journal} {\bibinfo  {journal} {Journal of Physics A: Mathematical and General}\ }\textbf {\bibinfo {volume} {17}},\ \bibinfo {pages} {L517} (\bibinfo {year} {1984})}\BibitemShut {NoStop}%
\bibitem [{\citenamefont {Burgess}\ \emph {et~al.}(2008)\citenamefont {Burgess}, \citenamefont {Holman},\ and\ \citenamefont {Hoover}}]{Burgess}%
  \BibitemOpen
  \bibfield  {author} {\bibinfo {author} {\bibfnamefont {C.~P.}\ \bibnamefont {Burgess}}, \bibinfo {author} {\bibfnamefont {R.}~\bibnamefont {Holman}},\ and\ \bibinfo {author} {\bibfnamefont {D.}~\bibnamefont {Hoover}},\ }\bibfield  {title} {\bibinfo {title} {Decoherence of inflationary primordial fluctuations},\ }\href@noop {} {\bibfield  {journal} {\bibinfo  {journal} {Physical Review D}\ }\textbf {\bibinfo {volume} {77}},\ \bibinfo {pages} {063534} (\bibinfo {year} {2008})}\BibitemShut {NoStop}%
\bibitem [{\citenamefont {Borghini}\ and\ \citenamefont {Gombeaud}(2012)}]{Borghini}%
  \BibitemOpen
  \bibfield  {author} {\bibinfo {author} {\bibfnamefont {N.}~\bibnamefont {Borghini}}\ and\ \bibinfo {author} {\bibfnamefont {C.}~\bibnamefont {Gombeaud}},\ }\bibfield  {title} {\bibinfo {title} {Heavy quarkonia in a medium as a quantum dissipative system: master-equation approach},\ }\href@noop {} {\bibfield  {journal} {\bibinfo  {journal} {European Physical Journal C}\ }\textbf {\bibinfo {volume} {72}},\ \bibinfo {pages} {2000} (\bibinfo {year} {2012})}\BibitemShut {NoStop}%
\bibitem [{\citenamefont {Baomar}\ \emph {et~al.}(2022)\citenamefont {Baomar} \emph {et~al.}}]{BAOMAR2022108266}%
  \BibitemOpen
  \bibfield  {author} {\bibinfo {author} {\bibfnamefont {H.}~\bibnamefont {Baomar}} \emph {et~al.},\ }\bibfield  {title} {\bibinfo {title} {{QTRAJ} 1.0: A lindblad equation solver for heavy-quarkonium dynamics},\ }\href@noop {} {\bibfield  {journal} {\bibinfo  {journal} {Computer Physics Communications}\ }\textbf {\bibinfo {volume} {273}},\ \bibinfo {pages} {108266} (\bibinfo {year} {2022})}\BibitemShut {NoStop}%
\bibitem [{\citenamefont {Yao}(2021)}]{Yao}%
  \BibitemOpen
  \bibfield  {author} {\bibinfo {author} {\bibfnamefont {X.}~\bibnamefont {Yao}},\ }\bibfield  {title} {\bibinfo {title} {Open quantum systems for quarkonia},\ }\href@noop {} {\bibfield  {journal} {\bibinfo  {journal} {International Journal of Modern Physics A}\ }\textbf {\bibinfo {volume} {36}} (\bibinfo {year} {2021})}\BibitemShut {NoStop}%
\bibitem [{\citenamefont {Penrose}(1996)}]{Penrose}%
  \BibitemOpen
  \bibfield  {author} {\bibinfo {author} {\bibfnamefont {R.}~\bibnamefont {Penrose}},\ }\bibfield  {title} {\bibinfo {title} {Gravity and state vector reduction},\ }\href@noop {} {\bibfield  {journal} {\bibinfo  {journal} {General Relativity and Gravitation}\ }\textbf {\bibinfo {volume} {28}},\ \bibinfo {pages} {581} (\bibinfo {year} {1996})}\BibitemShut {NoStop}%
\bibitem [{\citenamefont {Page}\ and\ \citenamefont {Wootters}(1983)}]{PageWootters1983}%
  \BibitemOpen
  \bibfield  {author} {\bibinfo {author} {\bibfnamefont {D.~N.}\ \bibnamefont {Page}}\ and\ \bibinfo {author} {\bibfnamefont {W.~K.}\ \bibnamefont {Wootters}},\ }\bibfield  {title} {\bibinfo {title} {Evolution without evolution: Dynamics described by stationary observables},\ }\href@noop {} {\bibfield  {journal} {\bibinfo  {journal} {Physical Review D}\ }\textbf {\bibinfo {volume} {27}},\ \bibinfo {pages} {2885} (\bibinfo {year} {1983})}\BibitemShut {NoStop}%
\bibitem [{\citenamefont {H{\"o}hn}\ \emph {et~al.}(2024)\citenamefont {H{\"o}hn}, \citenamefont {Russo},\ and\ \citenamefont {Smith}}]{Hohn2024PRD}%
  \BibitemOpen
  \bibfield  {author} {\bibinfo {author} {\bibfnamefont {P.~A.}\ \bibnamefont {H{\"o}hn}}, \bibinfo {author} {\bibfnamefont {A.}~\bibnamefont {Russo}},\ and\ \bibinfo {author} {\bibfnamefont {A.~R.~H.}\ \bibnamefont {Smith}},\ }\bibfield  {title} {\bibinfo {title} {Matter relative to quantum hypersurfaces},\ }\href@noop {} {\bibfield  {journal} {\bibinfo  {journal} {Physical Review D}\ }\textbf {\bibinfo {volume} {109}},\ \bibinfo {pages} {105011} (\bibinfo {year} {2024})}\BibitemShut {NoStop}%
\bibitem [{Note1()}]{Note1}%
  \BibitemOpen
  \bibinfo {note} {For strictly local GKLS generators, \(\Delta _{xy}\) reduces to variations induced by \(n^\mu \DOTSB \mapstochar \rightarrow n^\mu +\delta n^\mu \), and possible extrinsic-curvature corrections if present.}\BibitemShut {Stop}%
\bibitem [{\citenamefont {Kashiwagi}\ and\ \citenamefont {Matsumura}(2024)}]{KashiwagiMatsumuraPRA2024}%
  \BibitemOpen
  \bibfield  {author} {\bibinfo {author} {\bibfnamefont {K.}~\bibnamefont {Kashiwagi}}\ and\ \bibinfo {author} {\bibfnamefont {A.}~\bibnamefont {Matsumura}},\ }\bibfield  {title} {\bibinfo {title} {Markovian quantum master equation with {Poincar{\'e}} symmetry},\ }\href@noop {} {\bibfield  {journal} {\bibinfo  {journal} {Physical Review A}\ }\textbf {\bibinfo {volume} {109}},\ \bibinfo {pages} {052214} (\bibinfo {year} {2024})}\BibitemShut {NoStop}%
\bibitem [{\citenamefont {Chebotarev}\ and\ \citenamefont {Fagnola}(1998)}]{ChebotarevFagnola1998}%
  \BibitemOpen
  \bibfield  {author} {\bibinfo {author} {\bibfnamefont {A.~M.}\ \bibnamefont {Chebotarev}}\ and\ \bibinfo {author} {\bibfnamefont {F.}~\bibnamefont {Fagnola}},\ }\bibfield  {title} {\bibinfo {title} {Sufficient conditions for conservativity of quantum dynamical semigroups},\ }\href@noop {} {\bibfield  {journal} {\bibinfo  {journal} {Journal of Functional Analysis}\ }\textbf {\bibinfo {volume} {153}},\ \bibinfo {pages} {382} (\bibinfo {year} {1998})}\BibitemShut {NoStop}%
\bibitem [{\citenamefont {Oppenheim}(2023)}]{OppenheimPRX2023}%
  \BibitemOpen
  \bibfield  {author} {\bibinfo {author} {\bibfnamefont {J.}~\bibnamefont {Oppenheim}},\ }\bibfield  {title} {\bibinfo {title} {A postquantum theory of classical gravity?},\ }\href@noop {} {\bibfield  {journal} {\bibinfo  {journal} {Physical Review X}\ }\textbf {\bibinfo {volume} {13}},\ \bibinfo {pages} {041040} (\bibinfo {year} {2023})}\BibitemShut {NoStop}%
\bibitem [{\citenamefont {Oppenheim}\ \emph {et~al.}(2023)\citenamefont {Oppenheim}, \citenamefont {Sparaciari}, \citenamefont {{\v{S}}oda},\ and\ \citenamefont {Weller-Davies}}]{OppenheimNC2023}%
  \BibitemOpen
  \bibfield  {author} {\bibinfo {author} {\bibfnamefont {J.}~\bibnamefont {Oppenheim}}, \bibinfo {author} {\bibfnamefont {C.}~\bibnamefont {Sparaciari}}, \bibinfo {author} {\bibfnamefont {B.}~\bibnamefont {{\v{S}}oda}},\ and\ \bibinfo {author} {\bibfnamefont {Z.}~\bibnamefont {Weller-Davies}},\ }\bibfield  {title} {\bibinfo {title} {Gravitationally induced decoherence vs space-time diffusion},\ }\href@noop {} {\bibfield  {journal} {\bibinfo  {journal} {Nature Communications}\ }\textbf {\bibinfo {volume} {14}},\ \bibinfo {pages} {7910} (\bibinfo {year} {2023})}\BibitemShut {NoStop}%
\bibitem [{\citenamefont {Oppenheim}\ \emph {et~al.}(2024)\citenamefont {Oppenheim}, \citenamefont {Russo},\ and\ \citenamefont {Weller-Davies}}]{OppenheimPRD2024}%
  \BibitemOpen
  \bibfield  {author} {\bibinfo {author} {\bibfnamefont {J.}~\bibnamefont {Oppenheim}}, \bibinfo {author} {\bibfnamefont {A.}~\bibnamefont {Russo}},\ and\ \bibinfo {author} {\bibfnamefont {Z.}~\bibnamefont {Weller-Davies}},\ }\bibfield  {title} {\bibinfo {title} {Diffeomorphism invariant classical--quantum path integrals for {Nordstr{\"o}m} gravity},\ }\href@noop {} {\bibfield  {journal} {\bibinfo  {journal} {Physical Review D}\ }\textbf {\bibinfo {volume} {110}},\ \bibinfo {pages} {024007} (\bibinfo {year} {2024})}\BibitemShut {NoStop}%
\bibitem [{\citenamefont {Annby-Andersson}\ \emph {et~al.}(2022)\citenamefont {Annby-Andersson}, \citenamefont {Bakhshinezhad}, \citenamefont {Bhattacharyya}, \citenamefont {De~Sousa}, \citenamefont {Jarzynski}, \citenamefont {Samuelsson},\ and\ \citenamefont {Potts}}]{PhysRevLett.129.050401}%
  \BibitemOpen
  \bibfield  {author} {\bibinfo {author} {\bibfnamefont {B.}~\bibnamefont {Annby-Andersson}}, \bibinfo {author} {\bibfnamefont {F.}~\bibnamefont {Bakhshinezhad}}, \bibinfo {author} {\bibfnamefont {D.}~\bibnamefont {Bhattacharyya}}, \bibinfo {author} {\bibfnamefont {G.}~\bibnamefont {De~Sousa}}, \bibinfo {author} {\bibfnamefont {C.}~\bibnamefont {Jarzynski}}, \bibinfo {author} {\bibfnamefont {P.}~\bibnamefont {Samuelsson}},\ and\ \bibinfo {author} {\bibfnamefont {P.~P.}\ \bibnamefont {Potts}},\ }\bibfield  {title} {\bibinfo {title} {Quantum fokker-planck master equation for continuous feedback control},\ }\href {https://doi.org/10.1103/PhysRevLett.129.050401} {\bibfield  {journal} {\bibinfo  {journal} {Phys. Rev. Lett.}\ }\textbf {\bibinfo {volume} {129}},\ \bibinfo {pages} {050401} (\bibinfo {year} {2022})}\BibitemShut {NoStop}%
\end{thebibliography}%

\end{document}